**Frequency-Hopping Wave Engineering with Metasurfaces**

*Hiroki Takeshita, [1][†] Ashif Aminulloh Fathnan, [1][†] Daisuke Nita, [1][†] Shinya Sugiura [2] and Hiroki Wakatsuchi [1]\**

((Optional Dedication))

H. Takeshita, Dr. A. A. Fathnan, D. Nita, Prof. H. Wakatsuchi
[1]Department of Engineering, Nagoya Institute of Technology, Gokiso-cho, Showa, Nagoya, Aichi, 466-8555, Japan

Prof S. Sugiura
[2]Institute of Industrial Science, The University of Tokyo, Meguro, Tokyo, 153-8505, Japan

[†]These authors contributed equally to this work.
\*Corresponding author (wakatsuchi.hiroki@nitech.ac.jp)

**Abstract**
Wave phenomena can be artificially engineered by scattering from metasurfaces, which aids in the design of radio-frequency and optical devices for wireless communication, sensing, imaging, wireless power transfer and bio/medical applications. Scattering responses vary with changing frequency; conversely, they remain unchanged at a constant frequency, which has been a long-standing limitation in the design of devices leveraging wave scattering phenomena. Here, we present metasurfaces that can scatter incident waves according to two variables—the frequency and pulse width—in multiple bands. Significantly, these scattering profiles are characterized by how the frequencies are used in different time windows due to transient circuits. In particular, with coupled transient circuits, we demonstrate variable scattering profiles in response to unique frequency sequences, which can markedly increase the available frequency channels in accordance with a factorial function. Our proposed concept, which is analogous to frequency hopping in wireless communication, advances wave engineering in electromagnetics and related fields.

**1. Introduction**

Materials available in nature essentially exhibit frequency dispersions or varying electromagnetic characteristics in accordance with the frequency spectra of incident waves.[1] Such materials are usually represented by linear time-invariant (LTI) systems,[2,3] which ensures that the same scattering profiles are obtained whenever a constant frequency is input. However, these time-invariant scattering profiles limit the degrees of freedom for controlling electromagnetic waves, and incoming signals containing particular frequency components. From an engineering perspective, the assignment of frequency resources is maximized in accordance with the frequency bandwidth resolved.[4–6] However, if communication signals are distinguished depending on the frequency sequence, more degrees of freedom are obtained, increasing the channel capacity[7] (Fig. 1).

A key solution to overcoming the limitations associated with LTI systems lies in introducing nonlinearity,[8–12] which provides capabilities superior to those obtained from a simple combination of linear media. In particular, the design of electromagnetic properties and responses is facilitated by artificially engineered subwavelength structures such as frequency-selective surfaces (FSSs),[13] metamaterials[14,15] and metasurfaces.[16] For instance, the electromagnetic properties of metasurfaces no longer become time-invariant but rather are tunable if nonlinear media are used with electric or thermal stimulation[17,18] or optical pumping.[19] Alternatively, the use of nonlinear circuits[20–22] recently has led to the development of time-varying metasurfaces with advanced characteristics, including nonreciprocity,[23,24] harmonic generation[25,26] and spread spectrum capabilities.[27] However, in practice, these advanced time-varying electromagnetic properties require precise (symbol-level) synchronization with transmitting sources as well as external control systems or energy resources such as direct-current (DC) sources, which limits the applicability of time-varying metasurfaces. Therefore, we present a paradigm for electromagnetic scattering control using passive metasurfaces that are capable of overcoming the restriction imposed by LTI systems surpassing the limitations of conventional frequency selectivity and allowing us to engineer incident waves with frequency sequences. The proposed concept of obtaining distinct responses with variation in frequency sequences is akin to frequency hopping, the spread-spectrum modulation scheme used for Bluetooth,[28] and is utilized here as an analogue filter to spatially control electromagnetic waves.

**2. Theory**

In this study, the pulse width is set to 50 ns or longer in the range of a few GHz, which ensures that the pulse spectrum is almost the same as the oscillating frequency.[29] However, under these circumstances, the scattering response essentially remains constant (see Supplementary Note 1). To overcome this constraint and achieve additional degrees of freedom, nonlinearity is an important factor since its use in electromagnetic media provides performance better than that from a simple combination of linear characteristics.[8] In particular, we use the recently proposed waveform-selective metasurfaces.[29–33] These nonlinear metasurfaces can be used to vary scattering parameters in accordance with the time width of an incident pulse (Supplementary Note 2) since the intrinsic resonant mechanisms of the metasurfaces are coupled to the transient responses of DC circuits. More specifically, we design a metasurface to selectively transmit an incoming wave by using the supercell depicted in Fig. 2a. Our metasurface is composed of a dielectric substrate (Rogers3003, 1.5-mm thick) and a conducting plane that has a periodic array of rectangular slits (7 mm × 15 mm) with a periodicity of 18 mm (see "Simulation model" in the Methods section for details). Lumped capacitors $C_1$ and $C_2$ are used to connect the gaps across the slits, which results in the adjustment of the resonant frequency. In addition, the gaps are connected by sets of four diodes that serve as diode bridge rectifiers. Thus, the incoming waveform is converted from a sine wave to its modulus waveform (i.e., |sin|) within the diode bridges, generating an infinite set of frequency components. However, most of the rectified energy appears at zero frequency, as seen from the Fourier series expansions (Supplementary Note 2). Because of the dominant zero-frequency component, transient responses appear when resistors are paired with inductors inside diode bridges.[29] Specifically, an inductor exhibits a strong electromotive force that blocks incoming electric charges during an initial period (Fig. 2b). An analogy is seen in a classic DC circuit comprising a DC power supply, a resistor and an inductor that shows a similar electromotive force during an initial period. Therefore, the lumped circuit elements are decoupled with the conducting geometry of the slit structure so that the intrinsic resonant mechanism is maintained to effectively transmit an incident wave. However, the electromotive force of the lumped inductors is gradually reduced by the zero frequency of the rectified electric charges (as seen in the above classic DC circuit), which results in the disruption of the resonance of the metasurface in the steady state.

We further exploit such waveform-selective mechanisms at several frequency bands by imposing $C_1 \neq C_2$ within metasurface unit cells. Importantly, waveform selectivity is related to DC circuit systems (Fig. 2b) activated by different frequency sources and therefore independent of each other. By either retaining the independence of the waveform-selective transient responses or introducing interlocking mechanisms, we demonstrate a particular type of wave scattering named frequency-hopping wave engineering.

## 3. Results

### 3.1. Multiband Operation

First, we numerically designed and demonstrated a dual-band waveform-selective metasurface based on the slit structure shown in Fig. 2a (see "Simulation models" and "Simulation methods" in the Methods section). The simulated transmittance is plotted on the left of Fig. 2c, where $C_1$ and $C_2$ were set to 0.1 pF and 0.6 pF, respectively (see also "Definition of transmittances" in the Methods section). According to the simulation results, short pulses were more strongly transmitted than continuous waves (CWs) at 2.4 GHz and 3.8 GHz because of the aforementioned transient response. The right of Fig. 2c more clearly shows that the transmittance varied between approximately 0.1 and 0.9 at the two different frequencies. These large transmittances at the two frequencies were attributed to the presence of the two different types of unit cells. Fig. 2d supports the observation that the transmittance peaks seen in the dual-band model were consistent with the transmittance peaks obtained by the single-band models, where $C_1$ and $C_2$ were both set to the same value of either 0.1 pF or 0.6 pF. Below, we show that our metasurfaces provide additional degrees of freedom for engineering electromagnetic wave propagation. Supplementary Note 3 provides more information on the simulation model of Fig. 2.

**3.2. Scattering Based on Frequency Combination**

The concept of waveform selectivity can be extended to more than two frequency bands by constructing supercells with additional unit cells with different capacitance values. Fig. 3a shows a periodic supercell based on four unit cells using different capacitors. As seen in Fig. 2b, where a metasurface unit cell was related to a DC circuit, the supercell of Fig. 3a is associated with four independent DC circuits, each of which has a DC source activated by a different incident frequency. These circuit systems reduce the voltages across the inductors in the steady state, and thus, the structure in Fig. 3a is designed to show limited transient transmittance for a CW or a long duration waveform at any of the four resonant frequencies related to the unit cells (see also "Definition of transmittances" in the Methods section). However, if the incident frequency is regularly switched between these four frequencies, then the inductor voltages can possibly be restored to zero voltage or the initial condition to enhance transient transmittance again. Therefore, through the optimization of the transient response and recovery time, a large transient transmittance can potentially be obtained for a long waveform. This concept is schematically shown in the equivalent circuit model of Fig. 3b, where four DC circuits are equipped with either a DC source or one of the three short circuits, which are repeatedly swapped to enable the inductor voltage to recover to zero while other frequencies are in use (see $V_{L1}$ on the right of Fig. 3b). The numerical simulation results of the model in Fig. 3a are plotted in Fig. 3c and Fig. 3d for single-frequency source cases and switched-frequency source cases, respectively. In the former, the input frequency was fixed at 2.0, 2.5, 3.3 or 4.1 GHz. Under these circumstances, the transient transmittance was shown to be merely 0.05 or less since the circuit across the slit responding to the incident CW signal was short-circuited, which weakened the transmitting mechanism of the metasurface. However, the transient transmittance was enhanced by increasing the total number of frequencies available for switching at every 100 ns time step. A large transient transmittance was obtained despite the use of the same frequency components due to both the transient circuit response and the properly designed recovery time of the inductor circuits. This effect is more evident in Fig. 3e, where the entire pulse period (i.e., one set of four pulse durations) was equally distributed among the four frequencies and varied from 0.1 μs to 30 μs. According to this simulation result, the transmittance averaged over the entire pulse period was maximized at a few hundred nanoseconds and then gradually decreased with increasing pulse period. This phenomenon occurred as the transmittance approached the steady-state condition, requiring an extended duration for the restoration of the inductor voltage. Therefore, a moderately long pulse period is important for efficiently increasing the transmittance. The use of the proposed concept provides not only waveform selectivity at several frequency bands but also an additional parameter for designing electromagnetic scattering, even at the same frequencies, which surpasses the capabilities of conventional frequency selectivity or two-parameter schemes. Thus, our design provides additional degrees of freedom for engineering wave propagation.

We experimentally validated such a waveform-selective metasurface operating in multiple frequency bands using the measurement sample shown in Fig. 3f (see "Measurement samples" and "Measurement methods" in the Methods section). This sample comprised four sets of two different unit cells with/without additional capacitors to vary the operating frequency and was designed to operate at 2.58 GHz and 3.46 GHz. The measured transient transmittance for a single-frequency CW oscillating at only one of these frequencies was approximately 0.07 (Fig. 3g). However, with the switching of the oscillation frequency between these two frequencies, the transient transmittance increased, as shown in Fig. 3h. Importantly, owing to the limitation of the bandwidth of the rectangular waveguide used (see "Measurement methods" in the Methods section), this structure was designed to operate at only two frequencies, unlike the simulation model, which was designed for four frequencies. Therefore, the improvement in the measured transient transmittance was relatively limited compared with that in the simulated transient transmittance (green curve in Fig. 3h and blue curve in Fig. 3d). However, a comparable level of transient transmittance was observed when an additional 200 ns was used for the entire pulse period, which ensured that the pulse period in the measurement was the same as that in the simulation (blue curve in Fig. 3h). We expect the measured transient transmittance to further improve with an increasing number of unit cells or operating frequencies. These results support the theory that waveform selectivity is feasible, and more importantly, the transient transmittance increases with the use of waveform selectivity or the transient circuit response at several frequencies. Additional numerical and experimental results related to Fig. 3 are shown in Supplementary Note 4 and Supplementary Note 5, respectively.

**3.3. Scattering Based on Frequency Sequence**

The scattering concept extended by frequency combination depicted in Fig. 3 can demonstrate even more selectivity for engineering incoming signals containing the same frequencies if the scattering is determined by a particular sequence of frequencies. Such selective scattering may be achieved using the schematic shown in Fig. 4a. This circuit configuration is similar to the one shown in Fig. 3b (despite the minor difference in the number of DC circuit systems). Additional variable resistors are introduced to maintain or disrupt each transient response. For instance, if the variable resistors have a large resistance, then a transient circuit response is obtained, as seen earlier. However, if these resistances are small, the transient response disappears, signifying that the transmittance of the metasurface is lowered. The resistances are controlled by coordinating each circuit system or the corresponding frequency with one of the other frequencies, which leads to scattering control based on the frequency sequence. This concept is achieved using a junction-gate field-effect transistor (JFET), as shown in Fig. 4b. The inductor voltage $V_{L1}$ is applied to the gate of the JFET in the left circuit as the bias voltage. Depending on the inductor voltage (with or without the incident wave to activate the left DC circuit), the effective resistance $R_{JFET2}$ between the drain and source changes, which varies how the left circuit behaves when the DC source is moved to the left circuit (i.e., when the incident frequency component is changed). Similarly, the inductor voltage $V_{L2}$ of the left circuit is used to switch the state of the right circuit. Since such a structure involves complicated nonlinear phenomena and causes instability during simulation, we experimentally validated our design using the measurement sample shown in Fig. 4c (see "Measurement samples" and "Measurement methods" in the Methods section). We combined three 300-ns pulses, for which the oscillation frequencies were 2.5, 3.3 and 3.9 GHz. When the input frequency was changed in the sequence of 3.3, 3.9 and 2.5 GHz, the entire transient transmittance was enhanced most strongly, as seen in Fig. 4d (see frequency sequence #1). With other frequency sequences, some of the JFETs were not properly activated, which resulted in lower transient transmittance. Such selectivity based on the frequency sequence can be used for longer waveforms, as demonstrated in Fig. 4d. The transmittance gap in the steady state is more evident in the spectrograms shown in Fig. 4e and Fig. 4f. In these results, the output signal exhibits a consistent level of transmittance across all three frequencies if the incident signal is excited at frequency sequence #1 in Fig. 4d instead of frequency sequence #6. Note that the steady-state case had only two distinguished sequences since only three frequencies were used. Consequently, sequences #1, #2 and #3 demonstrated consistent high-level transmittance, while sequences #4, #5 and #6 displayed corresponding low-level transmittance. By increasing the number of frequencies used $N$, we can attain more degrees of freedom, specifically in accordance with $N!$ (i.e., the factorial of $N$) during the initial period and $(N-1)!$ in the steady state. These results support the theory that electromagnetic scattering or wave propagation can be engineered according to the sequence of frequencies, which cannot be realized by the conventional concept of frequency selectivity or existing LTI systems. Supplementary Note 6 summarizes more results related to Fig. 4.

**3.4. Potential application in wireless communications**

As an example of potential applications, the concept of our metasurfaces is leveraged in wireless communications where additional selectivities are made available in accordance with particular frequency sequences. As opposed to Fig. 2 to Fig. 4, where incident waves were excited at single frequencies, wireless communications utilize wider bandwidths in realistic modulation schemes.[28] The performance of our metasurface demonstrated in Fig. 5a to Fig. 5c was experimentally evaluated by using chirp signals that swept the oscillation frequency during each 300-ns time slot as shown in Fig. 5b (cf. Fig. 5a). In this measurement, we used frequency sequences #1 and #6 shown in Fig. 4d. Under this circumstance, the measurement results plotted in Fig. 5c demonstrate that selectivity based on frequency sequences still appeared even if the bandwidth of the incident signal was broadened. For instance, when the oscillation frequency was changed between ± 200 MHz (i.e., the bandwidth wider than that used for Bluetooth and Wi-Fi[28]), the transmittance difference between the two frequency sequences was maintained at 3 dB.

In addition, we conducted measurements on our metasurface using interference signals. Notably, in practical wireless communication scenarios, multiple frequency bands are often utilized simultaneously to enhance data transmission efficiency.[34] However, the concurrent use of neighbouring frequency bands can lead to significant interference with intended communication signals.[28] As depicted in Fig. 5d and Fig. 5e, we intentionally employed neighbouring frequencies ($f_4$ = 2.9 GHz and $f_5$ = 3.6 GHz) as interference signals. The interference-to-signal ratio (I/S) was then varied by changing the magnitude of the interference signals to observe the impact on the transient transmittance of the original signal. Remarkably, Fig. 5f demonstrates the metasurface's robustness against interference from neighbouring frequencies, as the interference signals were effectively suppressed while maintaining the transmittance of the original signal (see Supplementary Note 7 for detailed results). This resilience can be attributed to the metasurface's inherent characteristic of selectively allowing transmittance for specific intended narrow bands. While the narrow-band characteristics are advantageous in suppressing interference and improving communication performance, this holds true only when the signal bandwidth is narrower than the metasurface's bandwidth. In cases where the signal bandwidth exceeds the metasurface's bandwidth, design adjustments must be made to avoid deterioration of the overall communication performance.

Furthermore, we conducted experiments on our metasurface involving the transmission of realistic communication signals containing binary data. This analysis allows us to investigate the performance of the proposed metasurface in selectively transmitting a predetermined frequency sequence between a transmitter and receiver. Specifically, we examined the relationship between the bit error rate (BER) and the signal-to-noise ratio (S/N) when utilizing the metasurface for communication with binary phase shift keying (BPSK) modulation and frequency-hopping carriers. The measurement setup for this evaluation is depicted in Fig. 5g, where the incident signal was subjected to BPSK modulation (see "Modulation method" in the Methods section in detail). After passing through the metasurface, the signal was demodulated in the presence of additive white Gaussian noise (AWGN). The left panel of Fig. 5h represents the received BPSK signals without AWGN, while the right panel shows the received signals with AWGN. BER calculations at the receiver's end indicate that when the metasurface was utilized with the correct frequency sequence ($f_1$, $f_2$ and $f_3$ or sequence #1 in Fig. 4d), the BER remained low, indicating successful transmission. However, when the frequency sequence differed from the predetermined sequence (specifically, $f_3$, $f_2$ and $f_1$ or sequence #6 in Fig. 4d), the BER increased, indicating errors in data transmission. The difference between the BERs became more pronounced when the S/N increased, as depicted in Fig. 5i. For example, at S/N = 5 dB, the difference reached almost tenfold. Supplementary Note 7 provides further insights into how this difference in BER impacts the transmission of realistic image data. Consequently, the results depicted in Fig. 5 support the potential application of our metasurfaces in wireless communications, offering additional selectivity based on frequency sequences.

## 4. Discussion and Outlook

The proposed concept offers capabilities beyond those of conventional frequency selectivity approaches. In the literature, wave scattering has been controlled by both frequency and pulse width, which adds an additional degree of freedom to classic FSSs or metasurfaces.[29,32,35] In this study, we extended the waveform selectivity to more than one frequency band through multiband structures. Such multiband operation is observed in various fields, including metasurfaces and their applications.[1,13,36,37] However, unlike prior studies, our design provides more freedom to characterize electromagnetic properties/responses as well as their applied devices by properly designing the transient response of loaded circuits, their recovery time and coupling, which yields an additional selectivity based on the factorial number of frequencies available. In particular, our metasurfaces can be utilized for advanced spatial filters to design wireless communication environments and to more efficiently use limited frequency resources even at the same frequencies.[4–6,28] While several relevant efforts on nonlinear metasurfaces have reported advanced wave selectivities, such as angular memory[38] and pulse-width dependency,[32,35,39] none of the existing studies have exploited selectivity based on frequency sequences. Moreover, although our concept appears analogous to the principle of the frequency-hopping spread spectrum (FHSS), it is also conceivable to design metasurfaces with a phase-hopping technique to conform to the direct sequence spread spectrum (DSSS). In such a scenario, the metasurfaces would incorporate phase modifications while maintaining a consistent carrier frequency within a specific frequency band. This approach may offer enhanced immunity to narrow-band interference and facilitate simpler synchronization techniques, which are based on only single frequency operation. However, the metasurfaces need to be sensitive to phase variations to accurately replicate the pseudorandom phase coding in the transmitted signal, necessitating further investigation in future studies.

As an example of potential applications of our metasurfaces, we presented measurement results in Fig. 5 to show how our metasurfaces can be leveraged in wireless communications. Importantly, while our evaluation focused on one-dimensional systems for simplicity, metasurfaces are designed over two-dimensional surfaces in more practical wireless communication scenarios. This expanded design space offers increased degrees of freedom, enabling additional capabilities such as optical angular momentum and angular dependence.[38,40] In particular, as extensively explored in recent years, our metasurfaces can be utilized for beamforming when the metasurfaces exhibit a phase gradient pattern over two-dimensional surfaces to achieve anomalous reflection/transmission.[41] Additionally, in Supplementary Note 4 we showed that our metasurfaces are not limited to one-dimensional systems but can be extended to respond to free-space waves as a first step towards applications in realistic wireless communication environments. Despite the limited transmittance observed in the current metasurface design illustrated in Fig. 4d, we emphasize the availability of diverse strategies amendable to our metasurfaces for enhancing their transmittance, including optimizing the resonance's $Q$ factors. Supplementary Note 4 provides information on how the transmitting characteristics can be further improved, which helps our metasurfaces more efficiently work in specific application scenarios.

In recent years, metasurfaces have gained significant attention for sixth-generation (6G) mobile communication systems, particularly as simplified analogue repeaters for millimetre-wave communications.[42–44] This surge of interest in practical applications is a relatively recent development, although the extensive studies on metasurfaces have been conducted over the past two decades. Consistent with this trend, our proposed concept of analogue spatial frequency-hopping (frequency-sequence) filters offers the potential to transform the design approaches in future communication systems. One promising application of our metasurfaces is in physical-level information encryption, where they can be utilized within secure communication systems that leverage beamforming techniques. As described by the wiretap channel model,[45–47] our metasurfaces can increase the secrecy capacity by maximizing the beamforming gain towards the legitimate receiver while intentionally degrading the signal quality at the eavesdropper's location. Although our current proposed scheme does not directly create a beam or null for physical layer security purposes, further research can be conducted to realize beamforming techniques specifically tailored to the selectivity based on frequency sequences in our metasurfaces. Potential directions for future work could involve leveraging the periodicity of the proposed metasurfaces within the supercells to engineer diffraction into various angles, thereby enabling controlled beamforming. Another approach could involve utilizing a multilayer structure, with one layer acting as a beamformer while our metasurfaces remain in the other layer. Note that similar works have realized beamforming metasurfaces based on the incoming pulse width for cloaked antennas[39] and reconfigurable intelligent surfaces (RISs).[41]

We proposed a scattering paradigm that is not limited by the conventional concept of frequencies or characteristics of LTI systems. We theoretically presented a methodology to design metasurfaces loaded with circuit components such as diodes and numerically and experimentally validated the ability of these metasurfaces to distinguish different waves even at the same frequencies in accordance with their pulse widths. Our metasurfaces were designed to operate in more than one frequency band, which signifies that the scattering profile can be designed fully by using frequency and pulse width. Importantly, the metasurfaces enhanced transient transmittance if the incident frequency was changed (or hopped) repeatedly in particular combinations or sequences. Thus, our study introduces a new approach for designing electromagnetic materials[1,15] and related devices and systems[36,48] in wireless communications,[22,25] sensing,[49] imaging,[50] wireless power transfer[11] and bio/medical applications,[51] which are no longer bound by the number of frequencies but rather can be extended to a larger framework based on frequency-hopping wave engineering.

**Additional information**
**Supplementary information** is available for this paper.

**Data availability**

The data that support the findings of this study are available from the corresponding author upon request.

**Code availability**

The codes that are used to generate results in the paper are available from the corresponding author upon request.

References

1. Balanis, C. A. *Advanced engineering electromagnetics*. (John Wiley & Sons, 2012).

2. Siebert, W. M. *Circuits, signals, and systems*. (MIT press, 1986).

3. Antsaklis, P. J. & Michel, A. N. *Linear systems*. vol. 8 (Springer, 1997).

4. *Federal Communications Commission, United States of America. FCC ONLINE TABLE OF FREQUENCY ALLOCATIONS 47 C.F.R. § 2.106, Revised on July 1,* https://transition.fcc.gov/oet/spectrum/table/fcctable.pdf *(2022)*.

5. Electronic Communications Committee. The European Table of Frequency Allocations and Applications in The Frequency Range 8.3 KHz to 3000 GHz (ECA TABLE*)*. in *Proceedings of European Conference of Postal and Telecommunications Administrations* (Electronic Communications Committee, 2020).

6. The Ministry of Internal Affairs and Communications (MIC) Japan. *MIC Frequency Assignment Plan, September 2021*, https://www.tele.soumu.go.jp/e/adm/freq/search/share/plan.htm (2021).

7. Andrews, M. R., Mitra, P. P. & deCarvalho, R. Tripling the capacity of wireless communications using electromagnetic polarization. *Nature* **409**, 316–318 (2001).

8. Denz, C., Flach, S. & Kivshar, Y. S. *Nonlinearities in Periodic Structures and Metamaterials*. (Springer, 2010). doi:10.1007/978-3-642-02066-7.

9. Lapine, M., Shadrivov, I. V. & Kivshar, Y. S. Colloquium: Nonlinear metamaterials. *Rev Mod Phys* **86**, 1093–1123 (2014).

10. Lee, J. *et al.* Giant nonlinear response from plasmonic metasurfaces coupled to intersubband transitions. *Nature* **511**, 65–69 (2014).

11. Assawaworrarit, S., Yu, X. & Fan, S. Robust wireless power transfer using a nonlinear parity–time-symmetric circuit. *Nature* **546**, 387–390 (2017).

12. Liu, Z. *et al.* High-Q Quasibound states in the continuum for nonlinear metasurfaces. *Phys Rev Lett* **123**, 253901 (2019).

13. Munk, B. A. *Frequency selective surfaces: theory and design*. (John Wiley & Sons, 2005). doi:10.1002/0471723770.

14. Smith, D. R., Padilla, W. J., Vier, D. C., Nemat-Nasser, S. C. & Schultz, S. Composite medium with simultaneously negative permeability and permittivity. *Phys Rev Lett* **84**, 4184–4187 (2000).

15. Engheta, N. & Ziolkowski, R. W. *Metamaterials: physics and engineering explorations*. (John Wiley & Sons, 2006). doi:10.1002/0471784192.

16. Yu, N. *et al.* Light propagation with phase discontinuities: generalized laws of reflection and refraction. *Science* **334**, 333–337 (2011).

17. Chen, H.-T. *et al.* Active terahertz metamaterial devices. *Nature* **444**, 597–600 (2006).

18. Driscoll, T. *et al.* Memory metamaterials. *Science* **325**, 1518–1521 (2009).

19. Xiao, S. *et al.* Loss-free and active optical negative-index metamaterials. *Nature* **466**, 735–738 (2010).

20. Wu, Z., Ra'di, Y. & Grbic, A. Tunable metasurfaces: a polarization rotator design. *Phys Rev X* **9**, 011036 (2019).

21. Taravati, S. & Eleftheriades, G. V. Full-duplex reflective beamsteering metasurface featuring magnetless nonreciprocal amplification. *Nat Commun* **12**, 4414 (2021).

22. Zhang, L. *et al.* A wireless communication scheme based on space- and frequency-division multiplexing using digital metasurfaces. *Nat Electron* 1–10 (2021).

23. Nagulu, A., Reiskarimian, N. & Krishnaswamy, H. Non-reciprocal electronics based on temporal modulation. *Nat Electron* **3**, 241–250 (2020).

24. Wang, X., Díaz-Rubio, A., Li, H., Tretyakov, S. A. & Alù, A. Theory and design of multifunctional space-time metasurfaces. *Phys Rev Appl* **13**, 044040 (2020).

25. Zhang, L. *et al.* Space-time-coding digital metasurfaces. *Nat Commun* **9**, 4334 (2018).

26. Liu, M., Powell, D. A., Zarate, Y. & Shadrivov, I. V. Huygens' metadevices for parametric waves. *Phys Rev X* **8**, 031077 (2018).

27. Wang, X. & Caloz, C. Pseudorandom sequence (space) time-modulated metasurfaces: principles, operations, and applications. *IEEE Antennas Propag Mag* **64**, 135–144 (2022).

28. Goldsmith, A. *Wireless communications*. (Cambridge university press, 2005).

29. Wakatsuchi, H., Long, J. & Sievenpiper, D. F. Waveform selective surfaces. *Adv Funct Mater* **29**, 1806386 (2019).

30. Wakatsuchi, H., Kim, S., Rushton, J. J. & Sievenpiper, D. F. Waveform-dependent absorbing metasurfaces. *Phys Rev Lett* **111**, 245501 (2013).

31. Wakatsuchi, H. *et al.* Waveform selectivity at the same frequency. *Sci Rep* **5**, 9639 (2015).

32. Vellucci, S., Monti, A., Barbuto, M., Toscano, A. & Bilotti, F. Waveform-selective mantle cloaks for intelligent antennas. *IEEE Trans Antennas Propag* **68**, 1717–1725 (2020).

33. Imani, M. F. & Smith, D. R. Temporal microwave ghost imaging using a reconfigurable disordered cavity. *Appl Phys Lett* **116**, 054102 (2020).

34. Hanzo, L. L., Münster, M., Choi, B. J. & Keller, T. OFDM and MC-CDMA for broadband multi-user communications, WLANs and broadcasting. (2003) doi:10.1002/9780470861813.

35. Barbuto, M. *et al.* Waveguide components and aperture antennas with frequency- and time-domain selectivity properties. *IEEE Trans Antennas Propag* **68**, 7196–7201 (2020).

36. Balanis, C. A. *Antenna theory: analysis and design*. (John Wiley & Sons, 2016).

37. Watts, C. M., Liu, X. & Padilla, W. J. Metamaterial electromagnetic wave absorbers. *Adv Mater* **24**, OP98–OP120 (2012).

38. Valagiannopoulos, C., Sarsen, A. & Alù, A. Angular memory of photonic metasurfaces. *IEEE Trans Antennas Propag* **69**, 7720–7728 (2021).

39. Ushikoshi, D. *et al.* Pulse-driven self-reconfigurable meta-antennas. *Nat Commun* **14**, 633 (2023).

40. Xu, Y. *et al.* Reconfiguring structured light beams using nonlinear metasurfaces. *Opt. Express* **26**, 30930 (2018).

41. Fathnan, A. A. *et al.* Unsynchronized reconfigurable intelligent surfaces with pulse-width-based design. *IEEE Trans Veh Technol* **PP**, 1–6 (2023) (early access).

42. Basar, E. *et al.* Wireless communications through reconfigurable intelligent surfaces. *IEEE Access* **7**, 116753–116773 (2019).

43. Tang, W. *et al.* Wireless communications with programmable metasurface: new paradigms, opportunities, and challenges on transceiver design. *IEEE Wirel Commun* **27**, 180–187 (2020).

44. Shlezinger, N., Alexandropoulos, G. C., Imani, M. F., Eldar, Y. C. & Smith, D. R. Dynamic metasurface antennas for 6G extreme massive MIMO communications. *IEEE Wirel Commun* **28**, 106–113 (2021).

45. Wyner, A. D. The wire-tap channel. *Bell Syst. Tech. J.* **54**, 1355–1387 (1975).

46. Tsiftsis, T. A., Valagiannopoulos, C., Liu, H., Boulogeorgos, A.-A. A. & Miridakis, N. I. Metasurface-coated devices: a new paradigm for energy-efficient and secure 6G communications. *IEEE Veh Technol Mag.* **17**, 27–36 (2022).

47. Zheng, Y. *et al.* Metasurface‐assisted wireless communication with physical level information encryption. *Adv Sci* **9**, 2204558 (2022).

48. Ziolkowski, R. W., Jin, P. & Lin, C.-C. Metamaterial-inspired engineering of antennas. *Proc IEEE* **99**, 1720–1731 (2011).

49. Kaelberer, T., Fedotov, V. A., Papasimakis, N., Tsai, D. P. & Zheludev, N. I. Toroidal dipolar response in a metamaterial. *Science* **330**, 1510–1512 (2010).

50. Zheng, G. *et al.* Metasurface holograms reaching 80% efficiency. *Nat Nanotechnol* **10**, 308–312 (2015).

51. Li, Z., Tian, X., Qiu, C.-W. & Ho, J. S. Metasurfaces for bioelectronics and healthcare. *Nat Electron* **4**, 382–391 (2021).

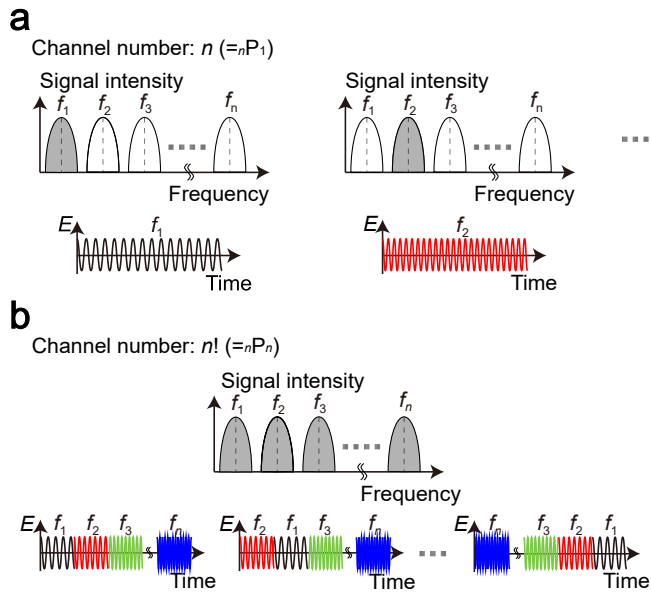

Fig. 1: Concept of the use of frequency-hopping metasurfaces to obtain additional degrees of freedom to engineer wave propagation. (a) Conventionally available frequency channels. The frequencies used are limited to the frequency resolution realized. (b) Frequency channels extended by the proposed concept, specifically by frequency-hopping wave engineering. Even with the same frequency resources, signals are preferentially distinguished if their frequency sequences are different.

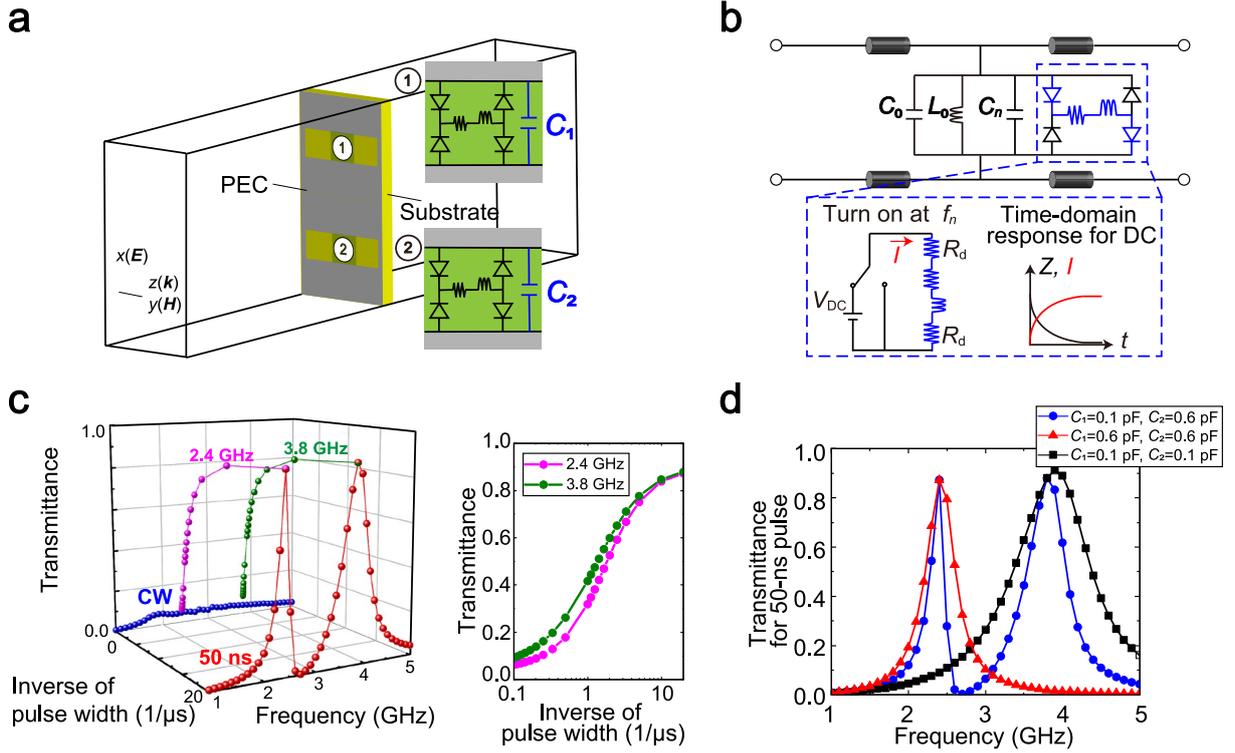

Fig. 2: Numerical demonstration of dual-band waveform-selective metasurfaces. (a) Supercell of the metasurface model based on the slit structure. $C_1$ and $C_2$ are used to adjust the operating frequencies of the unit cells. (b) Equivalent circuit model connected to the transmission line. The diode bridges across the slits and their internal circuit components (inside the top dashed box) can be approximated by a DC circuit, exhibiting a transient response even at the same frequency. (c) Scattering profiles (or transmittance profiles) extended from the classic frequency domain to the pulse-width domain. (d) Transmittance of 50-ns short pulses with $C_1 \neq C_2$ or $C_1 = C_2$.

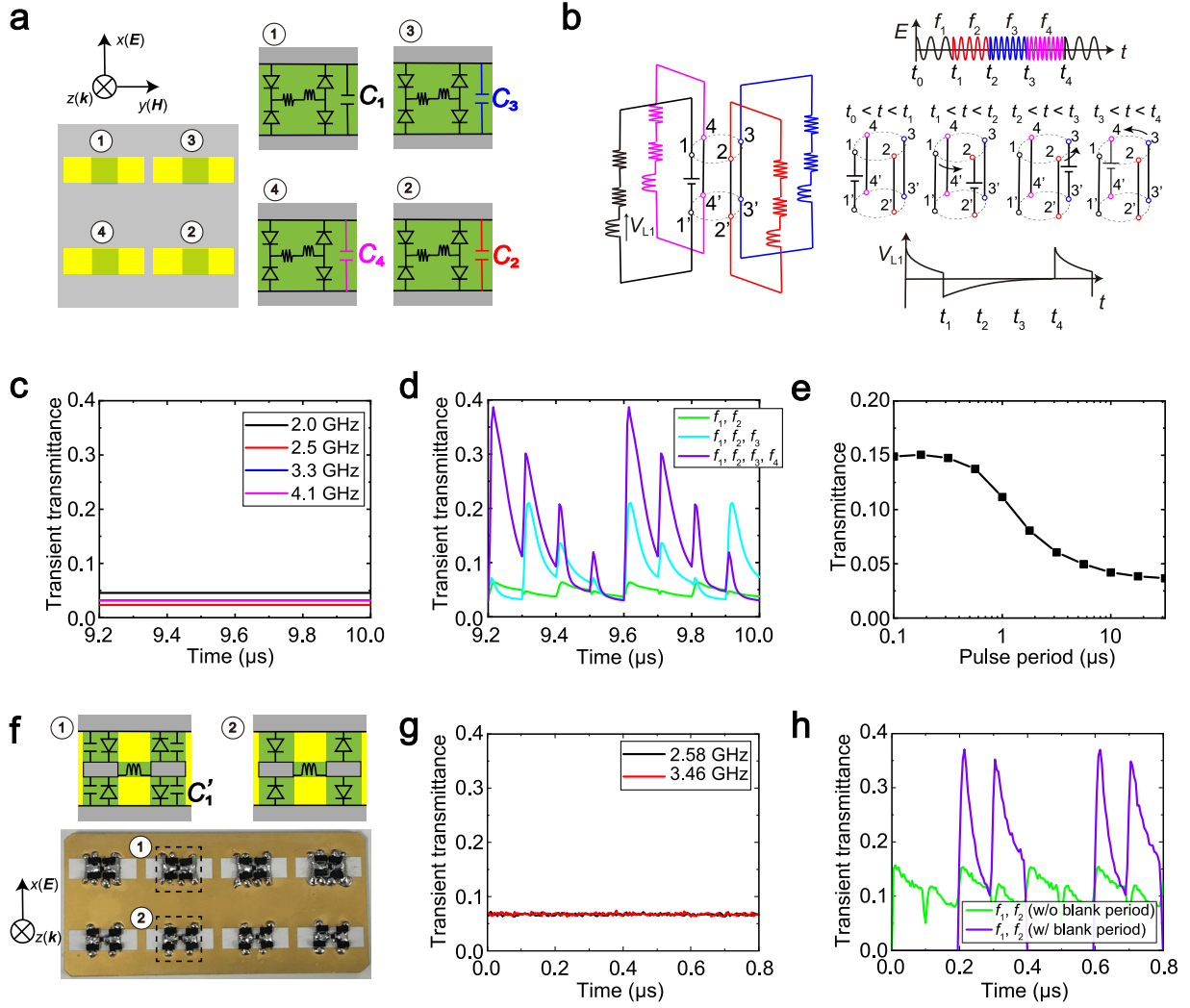

Fig. 3: Numerical and experimental demonstration of engineering wave propagation in accordance with frequency combinations. (a) Supercell of quadband waveform-selective metasurfaces. $C_1$, $C_2$, $C_3$ and $C_4$ are 0.1, 0.3, 0.6 and 1.1 pF, respectively. (b) Equivalent circuit system (left) and expected time-domain profiles (right). In the time domain, the incident frequency is repeatedly changed (top right), which corresponds to changing the position of the DC source (middle right). An inductor voltage can be restored to zero voltage while other frequencies are used (bottom right). (c) Simulated transient transmittance for single-frequency cases. (d) Simulated transient transmittance for switched-frequency cases. $f_1$, $f_2$, $f_3$ and $f_4$ are 2.0, 2.5, 3.3 and 4.1 GHz, respectively. (e) Average transmittance as a function of the entire pulse period. (f) Measurement sample designed to operate at two frequencies within a standard rectangular waveguide. (g) Measured transient transmittance for single-frequency cases. (h) Measured transient transmittance for the switched-frequency case. $f_1$ and $f_2$ are adjusted to 2.58 and 3.46 GHz, respectively.

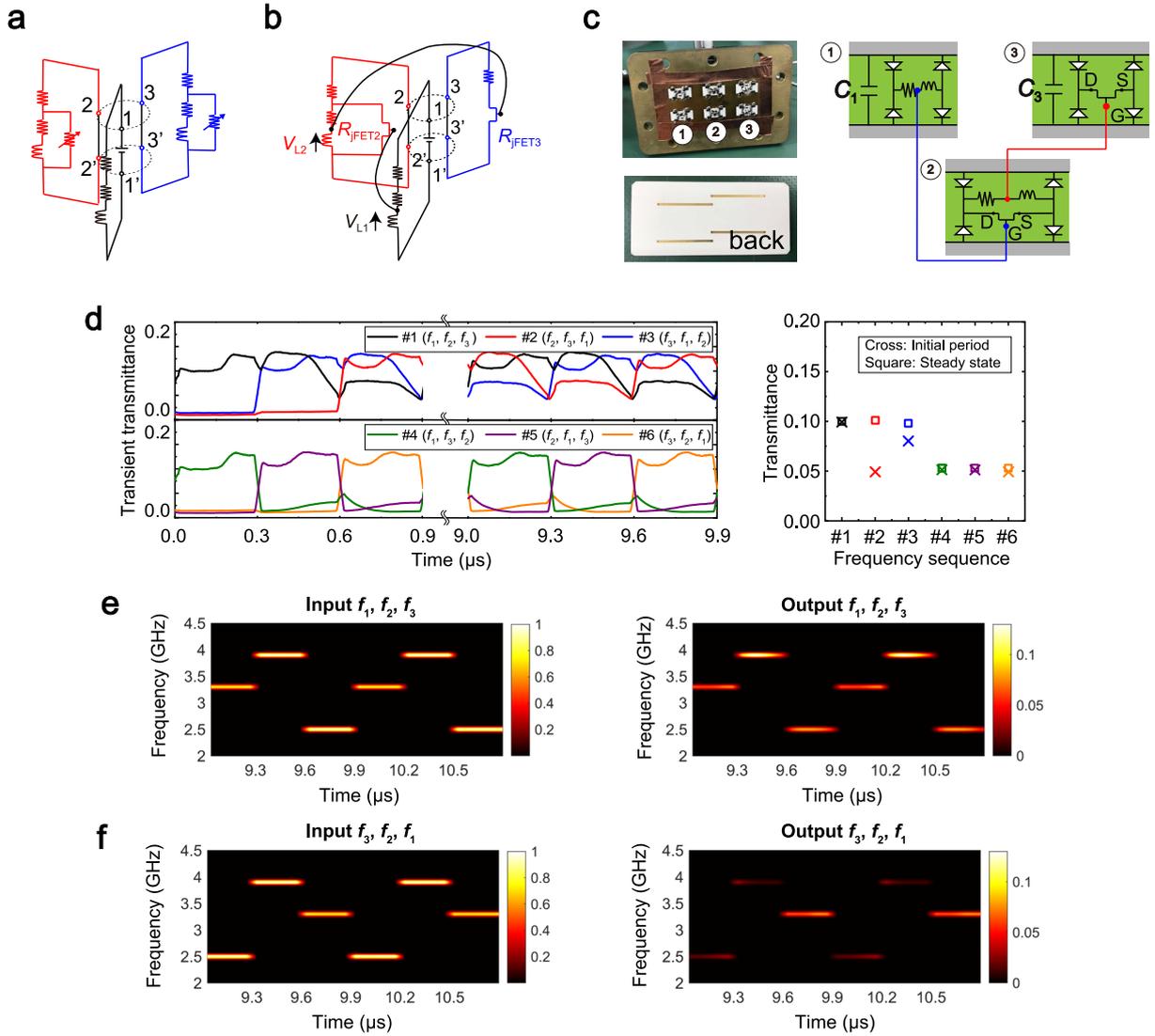

Fig. 4: Experimental demonstration of engineering wave propagation in accordance with frequency sequence. (a) Equivalent circuit system with variable resistors. The variable resistance values change in the time domain depending on the incoming frequency sequence. In the time domain, the incident frequency is repeatedly changed in a particular sequence, which determines how the position of the DC source is moved. (b) Specific circuit system. JFETs are included and biased by other circuits. (c) Measurement sample designed to operate at three frequencies. (d) Measured transient transmittances during the initial period and in the steady state (left) and their averages (right). $f_1$, $f_2$ and $f_3$ are 3.3, 3.9 and 2.5 GHz, respectively. During the initial period, the number of distinct frequency sequences is determined by $N!$, where $N$ represents the number of frequency channels available. In contrast, in the steady state, the number of distinct frequency sequences is reduced to a circular permutation of $N$, namely, $(N−1)!$. (e) Normalized spectrogram of the input and output signals (left and right, respectively) using the frequency sequence of $f_1$, $f_2$ and $f_3$. (f) Normalized spectrogram of the input and output signals (left and right, respectively) using the frequency sequence of $f_3$, $f_2$ and $f_1$. (e) and (f) correspond to sequences #1 and #6 in (d), respectively.

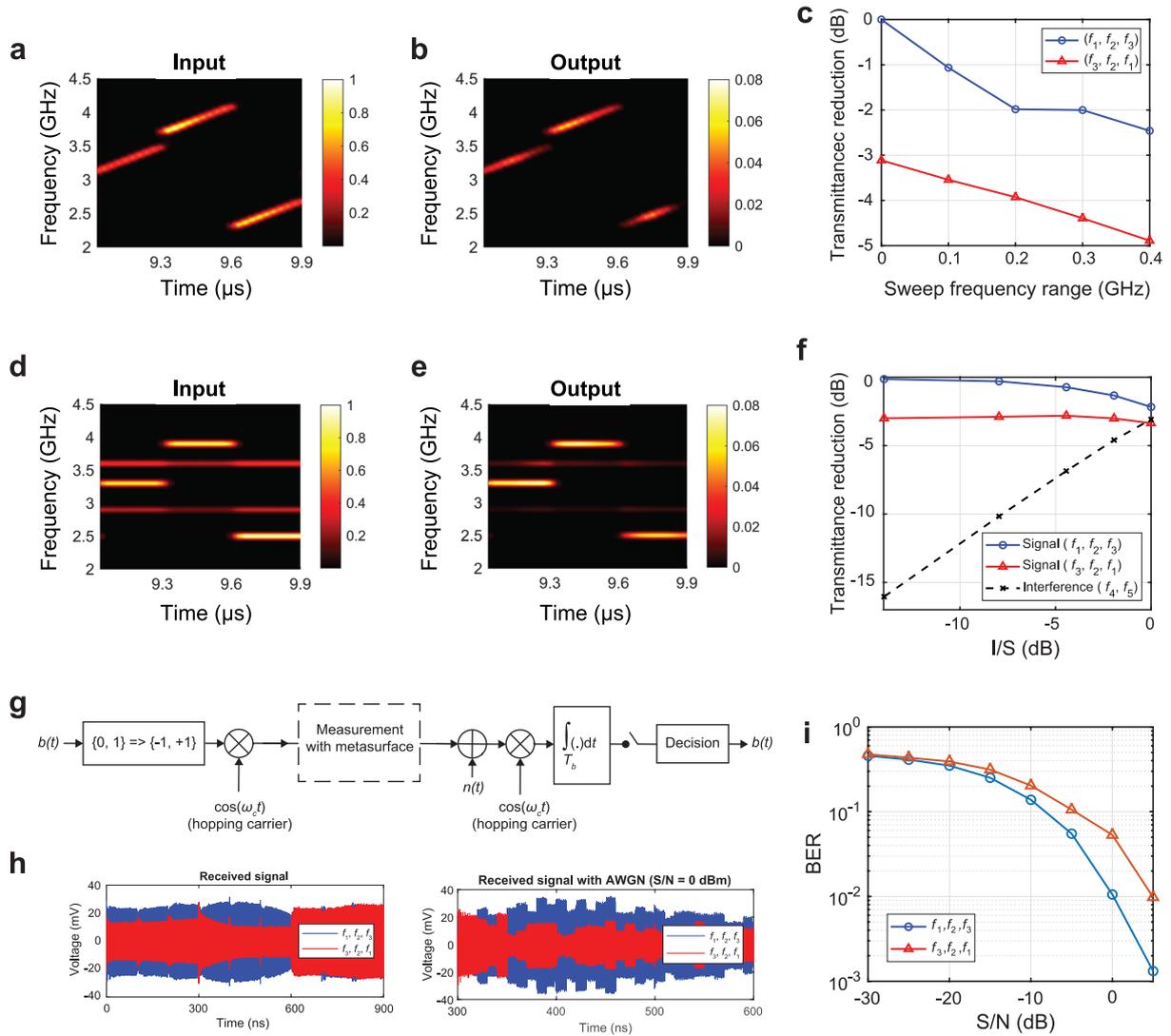

Fig. 5: Experimental demonstration of the metasurface towards realistic communication scenarios. (a, b) Spectrogram of (a) input and (b) output chirp signals sweeping the oscillation frequency during each 300-ns time slot. The centre frequency in each time slot hops between $f_1, f_2$ and $f_3$, which correspond to sequence #1 in Fig. 4d. In this example, the chirp signal is swept by 0.4 GHz in each time slot. (c) Transmittance reduction with different sweep frequency ranges. The transmitted energies of chirp signals using different frequency sequences are compared to the transmitted energy of sequence #1 shown in Fig. 4d. (d, e) Spectrogram of (d) input and (e) output frequency-sequence signals corresponding to sequence #1 in Fig. 4d with double-band interference signals at $f_4 = 2.9$ GHz and $f_5 = 3.6$ GHz. (f) Transmittance reduction with different interference-to-signal ratios (I/Ss). The transmitted energies of frequency-sequence signals and interference signals are compared to the transmitted energy of sequence #1 shown in Fig. 4d. Here, the interference signal magnitude ($f_4$ and $f_5$) is swept following the interference-to-signal ratio (I/S) from -14 dB to 0 dB. (g) Diagram block of communication scenario using the proposed metasurface and the BPSK modulation scheme. The time length for each bit $T_b$ is set to $T_b = 10$ ns, while the time slot for each frequency carrier is 300 ns (corresponding to 30 bits). (h) Examples of the received BPSK signals comparing two frequency sequences (left) before and (right) after implementing additive white Gaussian noise. (i) BER vs. S/N analysis using two different frequency sequences.

## 6. Methods

***Simulation models*:** Our simulation models were based on slit structures. Evenly spaced rectangular apertures were constructed in the conducting plate (perfect electric conductor: PEC) on a dielectric substrate (1.5-mm-thick Rogers3003) with periodic boundaries. In the default circuit configuration or an inductor-based circuit, each aperture was bridged by a set of four diodes (Broadcom HSMS286x series) where a resistor was connected to an inductor, as shown in Fig. 2 to Fig. 4. In Fig. 4, JFETs (Toshiba 2SK880-BL) were used to maintain or disrupt the transient response of the inductor circuit. The specific design dimensions and parameters are given in Supplementary Notes 3 to 6.

***Simulation method*:** We adopted a cosimulation method[29–31] to facilitate the design process of metasurfaces and optimized their performance by using the ANSYS Electronics Desktop Simulator (2020 R2). In this method, an electromagnetic metasurface model was first simulated in an electromagnetic simulator (HFSS). Lumped circuit components were replaced with lumped ports. These electromagnetic scattering profiles were used in a circuit simulator where lumped ports were connected to actual circuit components such as diodes, resistors and inductors. This was effectively the same as directly connecting the circuit components to the electromagnetic model via the lumped ports in electromagnetic simulations. However, the use of the cosimulation method markedly improved the simulation efficiency.

***Definition of transmittances*:** In this study we evaluated transmission characteristics using two terminologies, "(ordinary) transmittance" and "transient transmittance". The former transmittance was used in most simulations and measurements (e.g., Fig. 2c and Fig. 2d) to assess the frequency- and time-domain transmittances by dividing the total transmitted energy by the total incident energy. In contrast, the latter transmittance (i.e., transient transmittance) was used to more properly evaluate time-varying transmission characteristics in the time domain (e.g., in Fig. 3c and Fig. 3d) by dividing the moving average of the transmitted energy by that of the incident energy. However, when the entire performance of the time-varying characteristics was calculated (e.g., in Fig. 3e and the right panel of Fig. 4d), "(ordinary) transmittance" was still used.

*Measurement samples*: Measurement samples were designed based on the corresponding simulation models (see Supplementary Notes 5 and 6 for the dimensions and parameters used for the measurement samples). However, owing to the spatial limitations of the measurement setups, the number of unit cells was limited to six or eight. Circuit components were soldered to the conducting surfaces of the measurement samples.

*Measurement methods*: To evaluate the experimental performance of the measurement samples, frequency-domain characteristics were evaluated using a vector network analyser (VNA) (Keysight Technologies N5249A). The VNA was connected to coaxial cables and then a standard rectangular waveguide (WR284), where the measurement samples were fixed using copper tape. Time-domain results were obtained using an arbitrary waveform generator (AWG) (Keysight Technologies M8195A) and an oscilloscope (Keysight Technologies DSOX6002A).

*Modulation methods*: In addition to the above time-domain measurement methods, we evaluated the communication performance of the proposed metasurface. Here we used a system model involving the BPSK modulation scheme with a carrier frequency that hopped every 300 ns in the sequence of $f_1$ = 3.3 GHz, $f_2$ = 3.9 GHz and $f_3$ = 2.5 GHz or in the reversed sequence. A text file of the BPSK-modulated signal was first generated using MATLAB and input into the abovementioned AWG. The signal was sent to the rectangular waveguide (WR284) in which the metasurface was placed. AWGN was added to the transmitted signal to model a realistic noisy channel and conduct BER vs. S/N analysis. Fig. 5f shows the block diagram of the system model where $b(t)$ denotes the input binary data and $\omega_c$ is the angular frequency of the carrier signal (hopping between $f_1$, $f_2$ and $f_3$ with two distinct sequences). In the demodulation process, the received signal was multiplied by the same carrier frequency. An integrator and a threshold detector were used to remove harmonics and retrieve the corresponding bits, respectively.

**Acknowledgements**
This work was supported in part by the Japan Science and Technology Agency (JST) under Precursory Research for Embryonic Science and Technology (PRESTO) No. JPMJPR1933 and JPMJPR193A and under Fusion Oriented Research for Disruptive Science and Technology (FOREST), Japan Society for the Promotion of Science (JSPS) KAKENHI No. 17KK0114 and No. 21H01324 and the Japanese Ministry of Internal Affairs and Communications (MIC) under Strategic Information and Communications R&D Promotion Program (SCOPE) No. 192106007.

**Author contributions**
H.W. conceived of the idea and designed the project. H.T., A.A.F. and D.N. primarily performed simulations and measurements under the supervision of H.W. and S.S. All authors contributed to analyzing the results and editing the paper.

# Supplementary Information

**Frequency-Hopping Wave Engineering with Metasurfaces**

*Hiroki Takeshita, Ashif Aminulloh Fathnan, Daisuke Nita, Shinya Sugiura and Hiroki Wakatsuchi*

Supplementary Note 1: Frequency selectivity of conventional electromagnetic media
Supplementary Note 2: Frequency conversion by full wave rectification and time constants
Supplementary Note 3: Supplementary information of Fig. 2
Supplementary Note 4: Supplementary information of the simulations related to Fig. 3
Supplementary Note 5: Supplementary information of the measurements related to Fig. 3
Supplementary Note 6: Supplementary information of Fig. 4
Supplementary Note 7: Supplementary information of Fig. 5

**Supplementary Note 1: Frequency selectivity of conventional electromagnetic media**

In this Supplementary Note, we explain how metasurfaces are theoretically approximated by equivalent circuit models that are characterized by the frequency and that remain unchanged if the incoming frequency spectra are fixed. The characteristic impedance $Z$ of a metasurface may be approximated by a simple equivalent circuit model. For instance, the conducting geometry of the slit structure shown in Fig. 2a can be represented by a parallel circuit:

$$Z(\omega) = \left(\frac{1}{j\omega L} + j\omega C\right)^{-1}, \quad (1)$$

where $\omega$ is the angular frequency associated with frequency $f$ through $\omega = 2\pi f$, $j^2 = -1$ and $L$ and $C$ denote the entire inductive and capacitive components of the metasurface, respectively. Assuming a two-port network that includes the above structure, the reflection coefficient $S_{11}$ and transmission coefficient $S_{21}$ are obtained by

$$S_{11}(\omega) = \frac{Z(\omega) - Z_0}{Z(\omega) + Z_0}, \quad (2)$$

$$S_{21}(\omega) = \frac{Z(\omega) \cdot Z_0}{Z(\omega) + Z_0}, \quad (3)$$

where $Z_0$ represents the characteristic impedance of the medium next to the metasurface (vacuum in this study). Note that in Eqs. (1) to (3), the impedance of the dielectric substrate behind the conducting pattern is omitted for simplicity, although it can be represented by an additional transmission line. Moreover, the reflected energy $R_{PLS}$ and transmitted energy $T_{PLS}$ of the incident pulse are derived from[52]

$$R_{PLS}(\omega) = \frac{\int |b^-(\omega)|^2 d\omega}{\int |b^+(\omega)|^2 d\omega} = \frac{\int |b^+(\omega) S_{11}(\omega)|^2 d\omega}{\int |b^+(\omega)|^2 d\omega}, \quad (4)$$

$$T_{PLS}(\omega) = \frac{\int |b^+(\omega) S_{21}(\omega)|^2 d\omega}{\int |b^+(\omega)|^2 d\omega}, \quad (5)$$

where $b^+$ and $b^-$ are the magnitudes of the incident and reflected waves, respectively. Therefore, Eqs. (4) to (5) indicate that $R_{PLS}$ and $T_{PLS}$ remain the same if the frequency spectrum of the incident pulse is the same. Moreover, this study used pulsed sine waves, the electric field intensity $E$ of which followed the function below:[53]

$$E(\omega) = \frac{a}{2}\left(\frac{sin(\omega-\omega_0)T_{PW}}{0.5(\omega-\omega_0)} + \frac{sin(\omega+\omega_0)T_{PW}}{0.5(\omega+\omega_0)}\right), \tag{6}$$

where $\omega_0$ is the oscillating angular frequency and $a$ and $T_{PW}$ are the magnitude and pulse width of the incident pulse, respectively. In this study, the pulse width is set to 50 ns or longer in the range of a few GHz, which ensures that the pulse spectrum is almost the same as the oscillating frequency.[29] Hence, it is also clear from Eqs. (4) to (6) that under these circumstances, the scattering response essentially remains the same if the incoming frequency spectra remain unchanged.

## Supplementary Note 2: Frequency conversion by full wave rectification and time constants

This Supplementary Note explains how the diode bridge and the internal circuit elements of a unit cell of a waveform-selective metasurface can effectively be approximated by the DC circuit at the bottom of Fig. 2b.[54] First, due to the diode bridge, the waveform of an incident sine wave is converted to the waveform based on a modulus of the sine function. The frequency spectra of the rectified waveform can readily be obtained by using Fourier series expansion:

$$A(t) = \frac{a_0}{2} + \sum_{n=1}^{\infty} a_n \cos\left(\frac{2\pi nt}{T}\right) + \sum_{n=1}^{\infty} b_n \sin\left(\frac{2\pi nt}{T}\right), \quad (7)$$

where

$$a_n = \frac{2}{T}\int_0^T A(t) \cos\left(\frac{2\pi nt}{T}\right) dt, \quad (8)$$

$$b_n = \frac{2}{T}\int_0^T A(t) \sin\left(\frac{2\pi nt}{T}\right) dt, \quad (9)$$

$T$ and $n$ denote the period of the incident sine wave (i.e., $T = 1/f$) and the natural number, respectively. Suppose that the rectified waveform is a modulus of a sine function, namely,

$$A(t) = \left|\sin\left(\frac{2\pi t}{T}\right)\right|, \quad (10)$$

$a_n$ and $b_n$ are rearranged as follows:

$$a_n = \frac{4\cos\left(\frac{n\pi}{2}\right)^2 \cos(n\pi)}{(1-n^2)\pi}, \quad (11)$$

$$b_n = \frac{2(1+\cos(n\pi))\sin(n\pi)}{(n^2-1)\pi}. \quad (12)$$

According to Eqs. (11) and (12), the energy of the incident frequency $f$ is converted to those of other frequencies, as seen in Table S1. This table indicates that a large portion of the energy appears at zero frequency, which enables us to exploit the transient phenomena well known in DC circuits even if an alternating current (AC) signal comes in.

Next, as seen above, the internal circuit configuration of a unit cell of the waveform-selective metasurface is found to effectively respond to a DC signal. Under this assumption, it is well known that the transient voltage across the inductor $V_\text{L}$ can be readily estimated by $V_\text{L} = E_0 e^{-t/\tau_\text{L}}$, where $E_0$ and $t$ are the DC voltage applied and time, respectively, while the time constant $\tau_\text{L}$ is determined by the circuit components used[54]

$$\tau_L = L/(R_L + 2R_d). \tag{13}$$

In Eq. (13), $L$ and $R_\text{L}$ are the inductance and the resistance inside the diode bridges, respectively. $R_\text{d}$ represents the resistive component of the diodes used. Note that $R_\text{d}$ plays an important role in determining the time constant since $\tau_\text{L}$ becomes totally different without the use of $R_\text{d}$.[54] Eq. (13) indicates that the waveform-selective transient responses can be readily tailored by changing the circuit values.

Table S1: Energy at each frequency component after full wave rectification.

| Frequency component | Before rectification | After rectification |
|---|---|---|
| 0 | 0 | $(2/\pi)^2$ |
| $f$ | 1 | 0 |
| $2f$ | 0 | $(4/3\pi)^2$ |
| $3f$ | 0 | 0 |
| $4f$ | 0 | $(4/15\pi)^2$ |
| $5f$ | 0 | 0 |
| … | … | … |

**Supplementary Note 3: Supplementary information of Fig. 2**

This Supplementary Note provides information related to Fig. 2. First, the design parameters and circuit values used for Fig. 2 are summarized in Figure S1, Table S2, Table S3 and Table S4. While Fig. 2c depicts a dual-band waveform-selective metasurface, conventional single-band waveform-selective metasurfaces can be designed by using the same capacitance values for $C_1$ and $C_2$, as seen in Figure S2. This figure shows that the transmittance for a 50-ns short pulse is more enhanced than that for a CW in a single-frequency band that can be adjusted by the values of $C_1$ and $C_2$. The power dependence of the dual-band waveform-selective metasurface demonstrated in Fig. 2 is shown in Figure S3. This figure indicates that the metasurface is independent of the incoming waveform with a fixed frequency if the incident power is small enough (e.g., -20 dBm), as the diodes used are not turned on yet. However, by sufficiently increasing the input power to 10 dBm, the difference between the transmittance for a short pulse and that for a CW is maximized. Additionally, Figure S3 shows that with a further increment in the input power, the transmittance for a short pulse starts decreasing because the voltage across the diodes becomes larger than the breakdown voltage. Therefore, it is important to properly design both the turn-on voltage and the breakdown voltage of diodes.

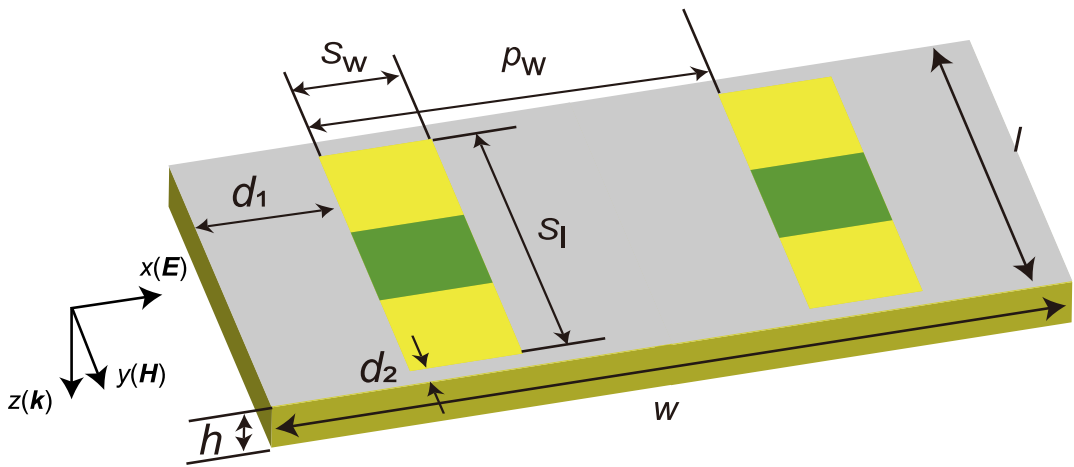

Figure S1: Supercell model used for Fig. 2. The design parameters and circuit parameters are shown in Table S2, Table S3 and Table S4. The substrate is Rogers3003.

Table S2: Design parameters used for Figure S1 (i.e., Fig. 2).

| Parameter | Length [mm] |
|---|---|
| $l$ | 18 |
| $w$ | 36 |
| $p_w$ | 18 |
| $s_l$ | 16 |
| $s_w$ | 5 |
| $h$ | 1.5 |
| $d_1$ | 6.5 |
| $d_2$ | 1 |

Table S3: Circuit values used for Figure S1 (i.e., Fig. 2). The value in parentheses represents a self-resonant frequency.

| Parameter | Value |
|---|---|
| $L$ | 1 mH (2.4 MHz) |
| $R_L$ | 10 Ω |
| $C_1$ | 0.1 pF |
| $C_2$ | 0.6 pF |

Table S4: SPICE parameters of diodes used for Figure S1 (i.e., Fig. 2).

| Parameter | Value |
|---|---|
| $I_{BV}$ | $1\times10^{-5}$ A |
| $I_s$ | $5\times10^{-8}$ A |
| $N$ | 1.08 |
| $R_s$ | 6 Ω |
| $C_j$ | 0.18 pF |
| $M$ | 0.5 |
| $P_B$ | 0.65 V |
| $P_r$ | 2 |
| $Ev$ | 0.69 eV |

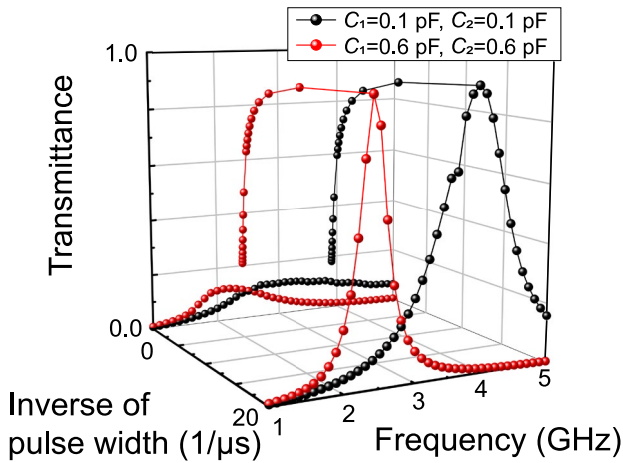

Figure S2: Transmittances of conventional single-band waveform-selective metasurfaces. $C_1$ and $C_2$ are set to the same values, as opposed to Fig. 2c, where the additional capacitances are different to achieve a dual-band operation.

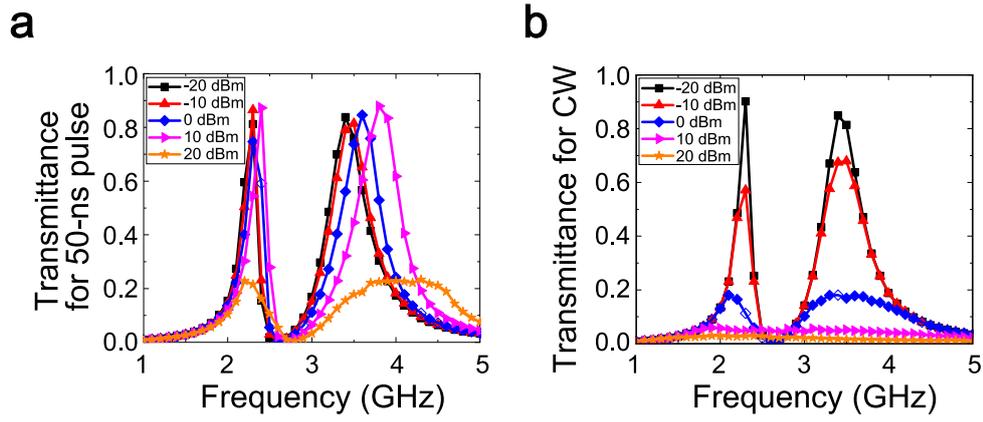

Figure S3: Power dependence of Fig. 2c. Transmittances for (a) 50-ns short pulses and (b) CWs.

**Supplementary Note 4: Supplementary information of the simulations related to Fig. 3**

This Supplementary Note provides information related to the numerical simulations depicted in Fig. 3. Additionally, supplementary information related to the measurement results of Fig. 3 is given in Supplementary Note 5. The design parameters and circuit values of the simulation model of Fig. 3 are the same as those used for Fig. 2 (see Figure S1, Table S2, Table S3 and Table S4). However, part of the circuit values is changed, as seen in Table S5. The frequency-domain profile of the simulation model in Fig. 3a is shown in Figure S4. As seen in this figure, the simulation model used in Fig. 3a exhibits four distinctive transmittance peaks at 2.0, 2.5, 3.3 and 4.3 GHz. However, to maximize the transient transmittance for an incoming signal whose oscillation frequency changes regularly, it may be necessary to further optimize the frequencies used. For instance, in Figure S5, 4.3 GHz (i.e., one of the four frequencies maximizing the frequency-domain transmittance) is changed between 4.1 and 4.4 GHz. The time-domain transient transmittance increases more at 4.1 GHz than at 4.3 GHz. This presumably occurs because with switching between four frequencies (2.0, 2.5, 3.3 and 4.3 GHz), which possibly leads to coupling between different cells, the voltage across diodes changes in the time domain compared to that in the frequency-domain result. Therefore, the related resistive component within the diodes is influenced, which varies the time constant and the recovery time of the entire circuit.

Table S5: Circuit values used for Fig. 3a.

| Parameter | Value |
|---|---|
| $L$ | 100 μH |
| $R_L$ | 10 Ω |
| $C_1$ | 0.1 pF |
| $C_2$ | 0.3 pF |
| $C_3$ | 0.6 pF |
| $C_4$ | 1.1 pF |

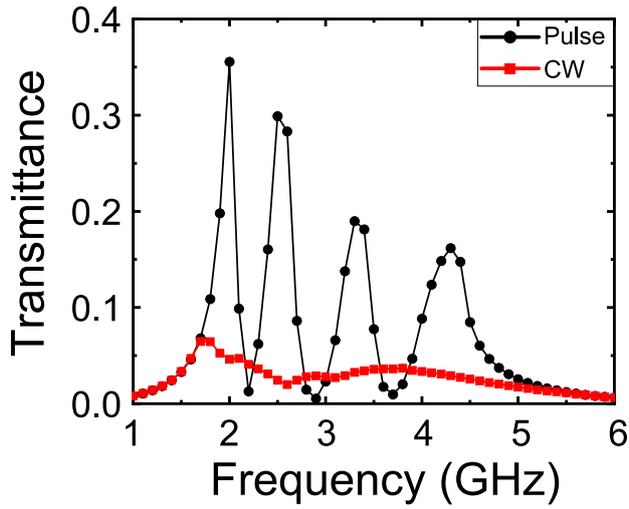

Figure S4: Frequency-domain profile of the simulation model of Fig. 3a. The pulse width and input power are set to 100 ns and 10 dBm, respectively.

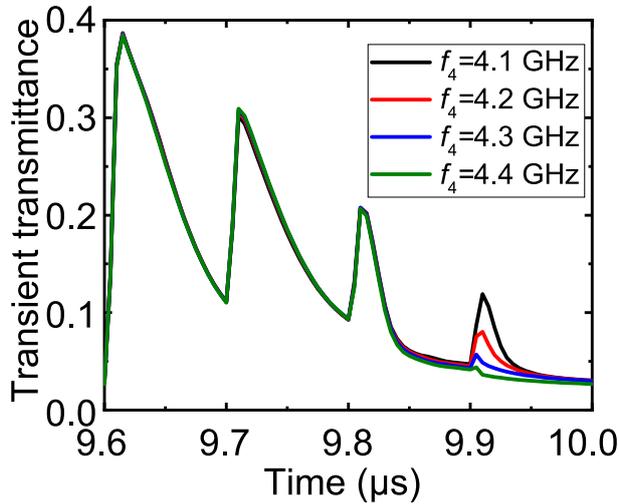

Figure S5: Simulated transient transmittance of Fig. 3a using various frequencies for $f_4$. Compared to Fig. 3d, the fourth frequency used is changed between 4.1 GHz and 4.4 GHz. Note that in Figure S4, the fourth transmittance peak does not appear at 4.1 GHz but at 4.3 GHz.

Regarding the recovery time to restore the inductor voltage, more information is provided in Figure S6. As seen in Figure S6a, we consider only one unit cell with periodic boundaries to readily evaluate the relationship between the recovery time and the circuit values used within the diode bridge. With an increase in the inductance from 10 μH to 10 mH, the transient transmittance for 100-ns pulses is enhanced near 4.9 GHz (Figure S6b and Figure S6c). This is because when a large inductance value is applied, the electromotive force of the inductor is maintained for short pulses, which leads to the intrinsic resonant mechanism of the slit structure being maintained. However, the inductor voltage is found to be slowly restored to zero voltage if a large inductance is used (see Figure S6d for the result at 4.9 GHz). Therefore, there is a trade-off between the recovery time and transient transmittance level in terms of the design of the inductance value. Moreover, the frequency-domain transmittance profile is also related to the resistor deployed within the diode bridge (Figure S6e). Specifically, the transient transmittance decreases with a reduction in the resistance, as the related time constant is also increased (see Figure S6f and Supplementary Note 2). In the time domain, the smaller the resistance is, the longer the recovery time becomes (Figure S6g). Importantly, however, increasing the resistance value reduces the difference between the transient transmittance during the initial period and that in the steady state (Figure S6f); thus, the waveform-selective characteristics disappear. Therefore, there is another trade-off between the transient transmittance level and waveform-selective performance in terms of the design of the resistance value. Based on these relationships, one may need to optimize the circuit values used for waveform-selective metasurfaces.

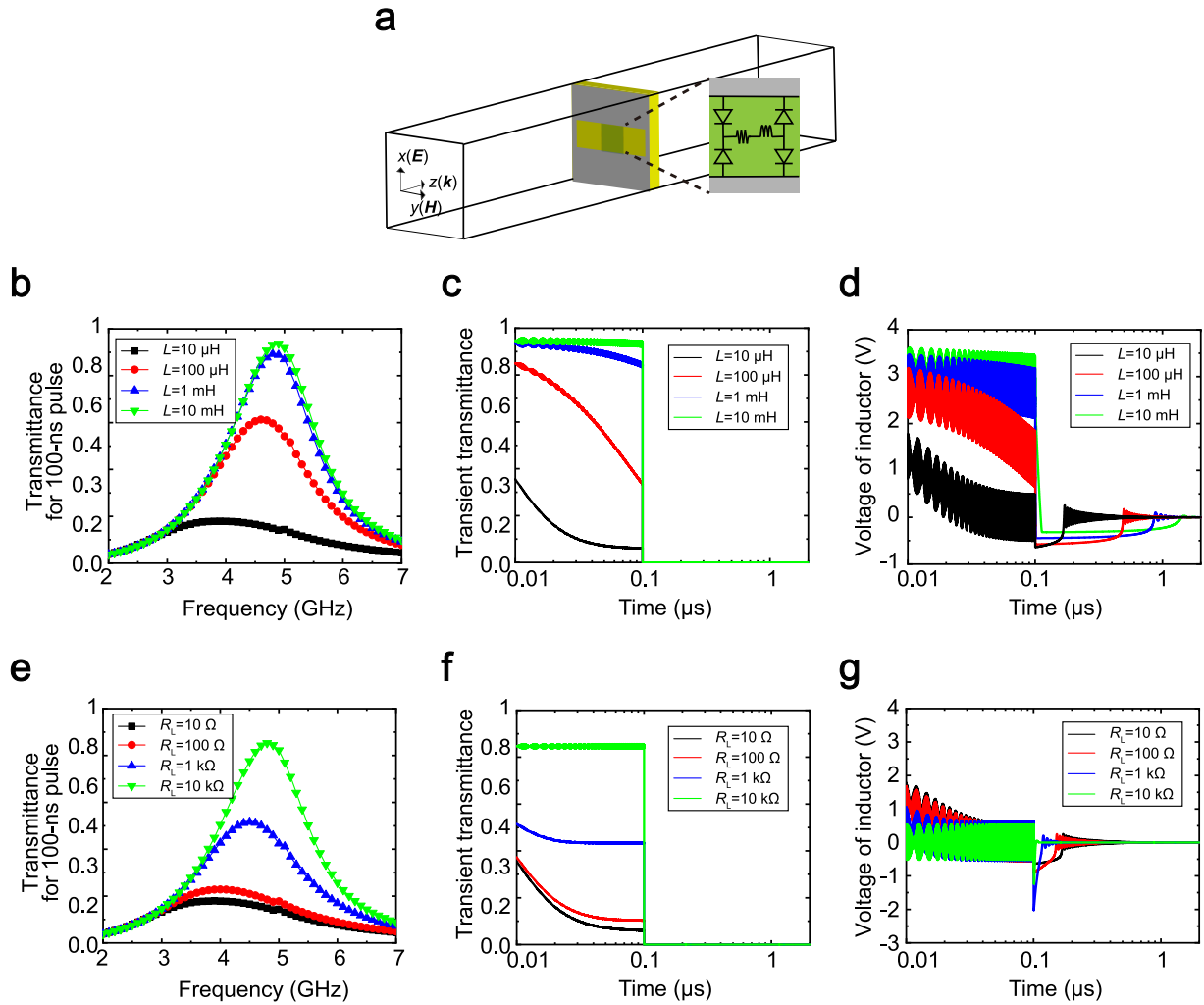

Figure S6: Relationship between circuit values and recovery time to restore inductor voltage. (a) Simulation model. Basically, the design parameters and circuit values are the same as those applied to Figure S1 (i.e., those applied to Fig. 2). (b) Frequency-domain profile with various inductances and 10-dBm input power and (c, d) corresponding time-domain profiles at 4.9 GHz. $R_L$ is fixed at 10 Ω. (e) Frequency-domain profile with various resistances and 10-dBm input power and (f, g) corresponding time-domain profiles. $L$ is fixed at 10 µH. In the time-domain results, incident waves are excited until 100 ns as 100-ns short pulses.

To better improve the overall transient transmittance of the simulation model associated with Fig. 3d, one may also optimize the oscillation duration of each frequency component. For instance, the oscillation duration is changed from 50 ns to 200 ns, as shown in Figure S7. As a result, the entire transient transmittance is found to change between 0.105 (#3 in Figure S7) and 0.140 (#1), compared to 0.138 in Fig. 3d. Note that in Figure S7, the transient transmittance for some individual frequencies is shown to be relatively large, although the average transmittance for the entire pulse period is not necessarily good (e.g., #3 in Figure S7). This is because the transient transmittance for other individual frequencies is lowered, which indicates that there is a trade-off between the transient transmittances of operating frequencies. Additionally, this transient transmittance change is strongly related to the recovery time and transient transmittance level determined by the circuit values used (see Figure S7 and the above discussion).

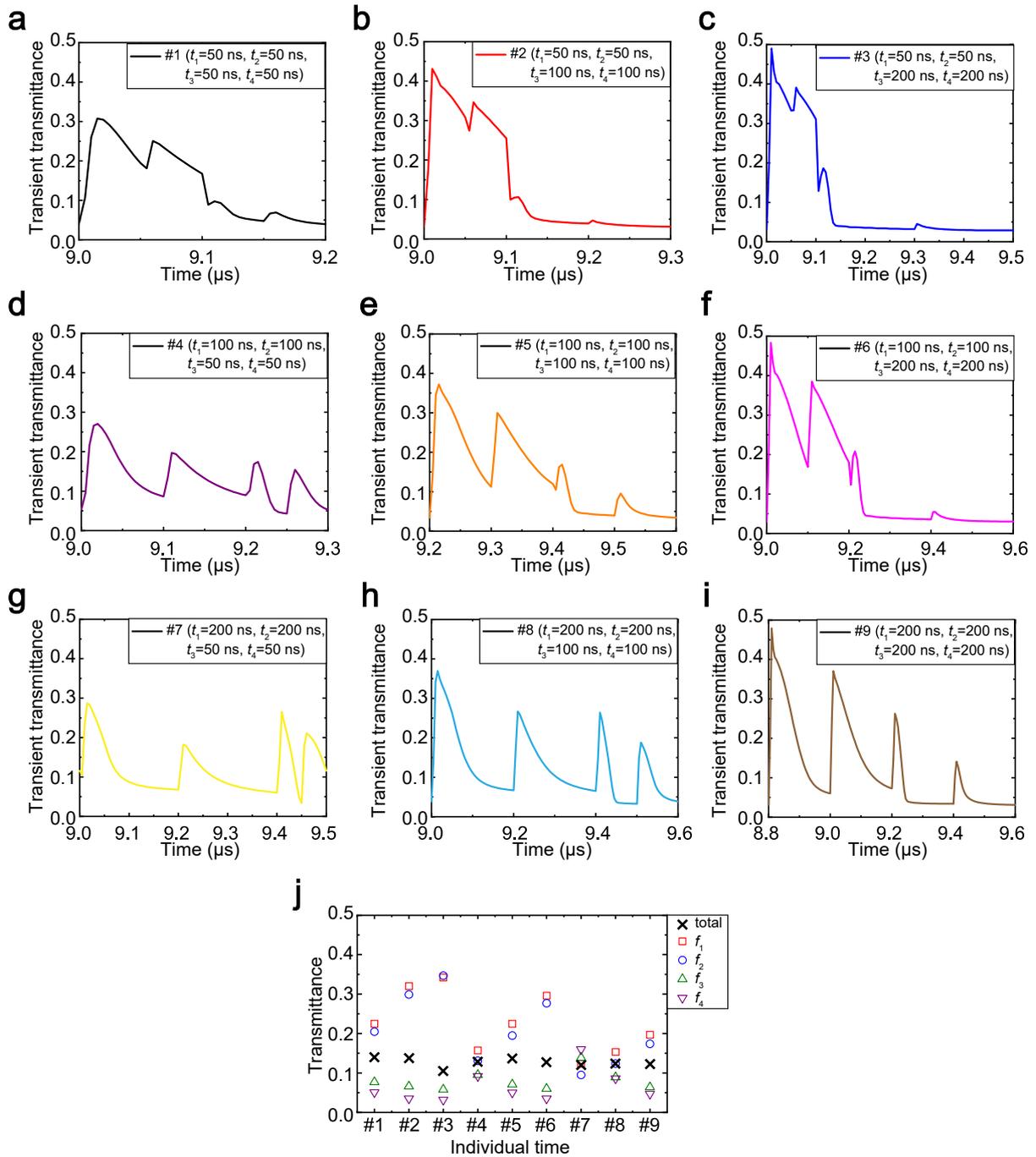

Figure S7: The transient transmittance of the simulation model depicted in Fig. 3a with various pulse duration scenarios. (a-i) From #1 to #9, the oscillation duration of each frequency is changed from 50 ns to 200 ns. (j) Average transmittance.

In contrast to Fig. 3, Fig. 4 shows transient transmittance varying in response to a particular frequency sequence. Such selectivity is not obtained by the simulation model of Fig. 3, as individual unit cells are not coupled to activate circuit structures of other cells. This is shown in Figure S8, where the frequency sequence of an incoming wave is changed. Figure S8 shows only a minor difference among different frequency sequences, which results from small coupling between unit cells. Note that this variation is much smaller than the one seen in Fig. 4d.

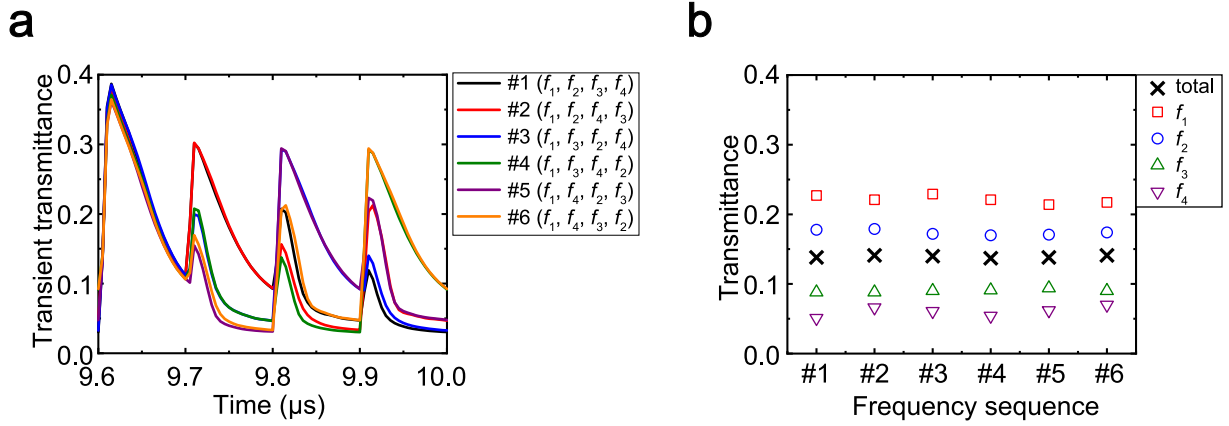

Figure S8: Transient transmittances of the simulation model drawn in Fig. 3a with different frequency sequence scenarios. (a) From #1 to #6, the frequency sequence used is changed. (b) Average transmittance.

Additionally, we numerically tested the quadband waveform-selective metasurface of Fig. 3a in free space to show that the use of our metasurfaces is not limited to one-dimensional systems. In this simulation we used a pair of standard horn antennas (Schwarzbeck BBHA 9120 D) as a transmitter and a receiver, as shown in Figure S9a and Figure S9b. The metasurface tested was composed of 6 × 6 unit cells (i.e., 3 × 3 supercells) and surrounded by a PEC wall to ensure that the incident wave transmitted through the metasurface. Other design parameters are given in Table S6. Under this circumstance, the metasurface continued to exhibit enhanced transient transmittance if the incident frequency was regularly switched, as seen in Figure S9c. This simulation result supports that the concept of our metasurfaces can be exploited in free space as well, which is important to find applications in, for instance, wireless power transfer and communications.

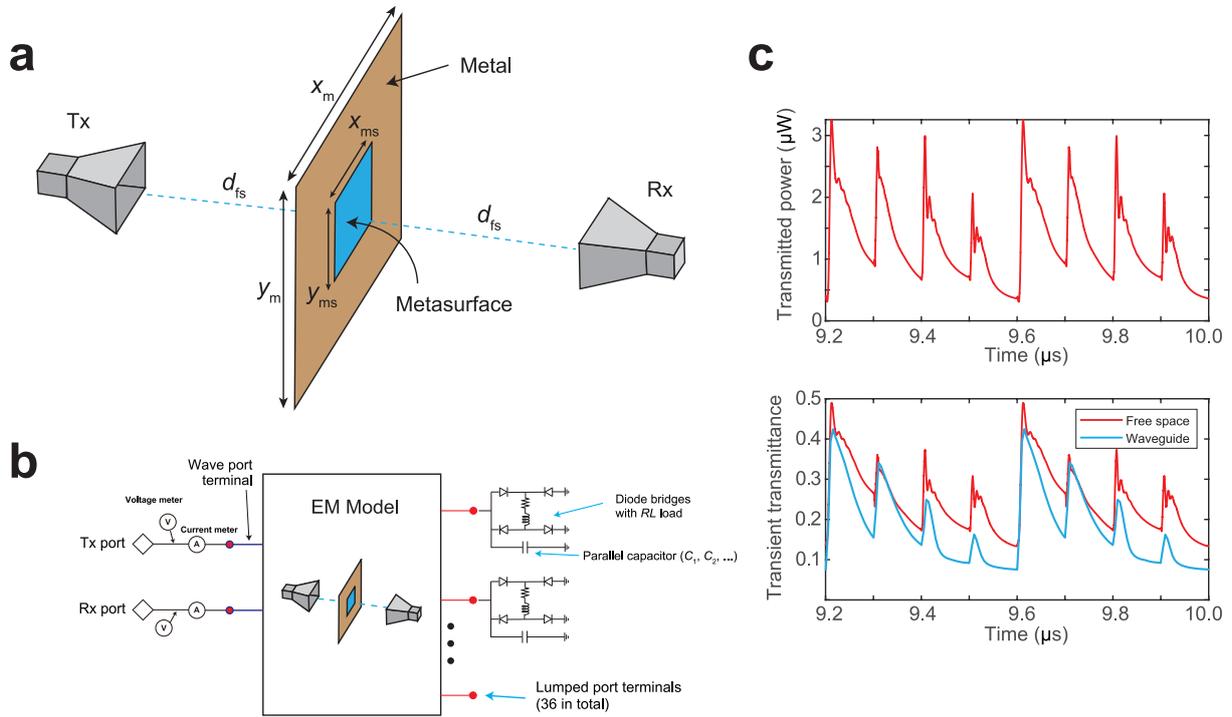

Figure S9: Simulation of the metasurface of Fig. 3a using the free-space configuration. (a) A pair of two horn antennas are used as a transmitter and a receiver in the electromagnetic simulation. The metasurface is deployed in a metallic conductor plate to narrow the metasurface area (lower the computational resources needed) while preventing the direction propagation of the signal between the transmitter and the receiver. The design parameters are shown in Table S6. (b) Circuit schematic integrating the electromagnetic simulation result. Two terminals are connected to the transmitter and receiver ports, and the remaining terminals are connected to diode bridges including inductors and resistors with parallel capacitors. (c) Simulation results of the cosimulation method. The top panel shows the transmitted power at the receiver when the transmitter generates a signal of 20 dBm. The bottom panel shows a transient transmittance that is obtained after normalization to the transmitted power when the metasurface is absent. Comparison to the waveguide simulation (the blue line and the same as Fig. 3d) shows a good agreement between both configurations, indicating the possibility of extending the proposed metasurface to free-space application scenarios.

Table S6: Design parameters used for Figure S9.

| Parameter | Length [mm] |
|---|---|
| $x_{ms}$ | 108 |
| $y_{ms}$ | 108 |
| $x_m$ | 420 |
| $y_m$ | 320 |
| $d_{fs}$ | 156 |

Moreover, compared to the dual-band waveform-selective metasurface results (e.g., in Fig. 2c), the level of time-domain transient transmittance was relatively lowered in Fig. 3. This reduced performance results from multiple factors including loss in substrate and circuit elements. However, the transient transmittance can be improved by adjusting the $Q$ factor. By readily demonstrating the relationship between the transient transmittance and the $Q$ factor, we performed simplified circuit simulations where our metasurface shown in Fig. 3a was approximated by lumped circuit elements that were connected to transmission lines. First, as seen in Figure S10a, the transmitting profiles of a single slit were well represented by a capacitor $C_0$ paired with an inductor $L_0$ in parallel as well as a diode bridge including an inductor connected to a resistor in series. Here, to represent the inductive and resistive components contributed by the dielectric substrate, $L_{add}$ and $R_{add}$ were connected to capacitor $C_0$ in parallel and series, respectively. By varying the capacitors $C_0$ and $C_1$ and inductor $L_0$, we vary the $Q$ factor of the resonance, as detailed in Table S7. Simulation results from the linear network analysis in Figure S10c illustrated that in the absence of any parasitic resistance ($R_{add}$ = 0 Ω), the transmittance remained consistently high at the resonance frequency, regardless of the $Q$ factor values. Nevertheless, when incorporating a minor amount of parasitic resistance ($R_{add}$ = 3 Ω), the transmittance exhibits variations depending on the specific value of the $Q$ factor. Here, the transmittance was markedly improved if the $Q$ factor was relatively lowered as seen by the yellow line in Figure S10c. In Figure S10d, the structure was extended to have four unit cells and realize quadband operation, and similar transmittance profiles were observed. Under this situation, similar to the single resonance scenario, when there was no parasitic resistance ($R_{add}$ = 0 Ω), the maximum transmittance consistently remained high across different $Q$ factor values, as observed in Figure S10e. However, in the case of introducing a small parasitic resistance ($R_{add}$ = 3 Ω), the transmittance exhibited variations corresponding to the $Q$ factor, as depicted in Figure S10f. In Figure S10g and Figure S10h, the incident frequency was regularly changed to obtain transient transmittance similar to Fig. 3d, in which the utilization of a lower $Q$ factor is more favourable for a higher transient transmittance. Therefore, the transient transmittance can potentially be improved by minimizing the lossy elements of the metasurfaces and lowering the resonance's $Q$ factor.

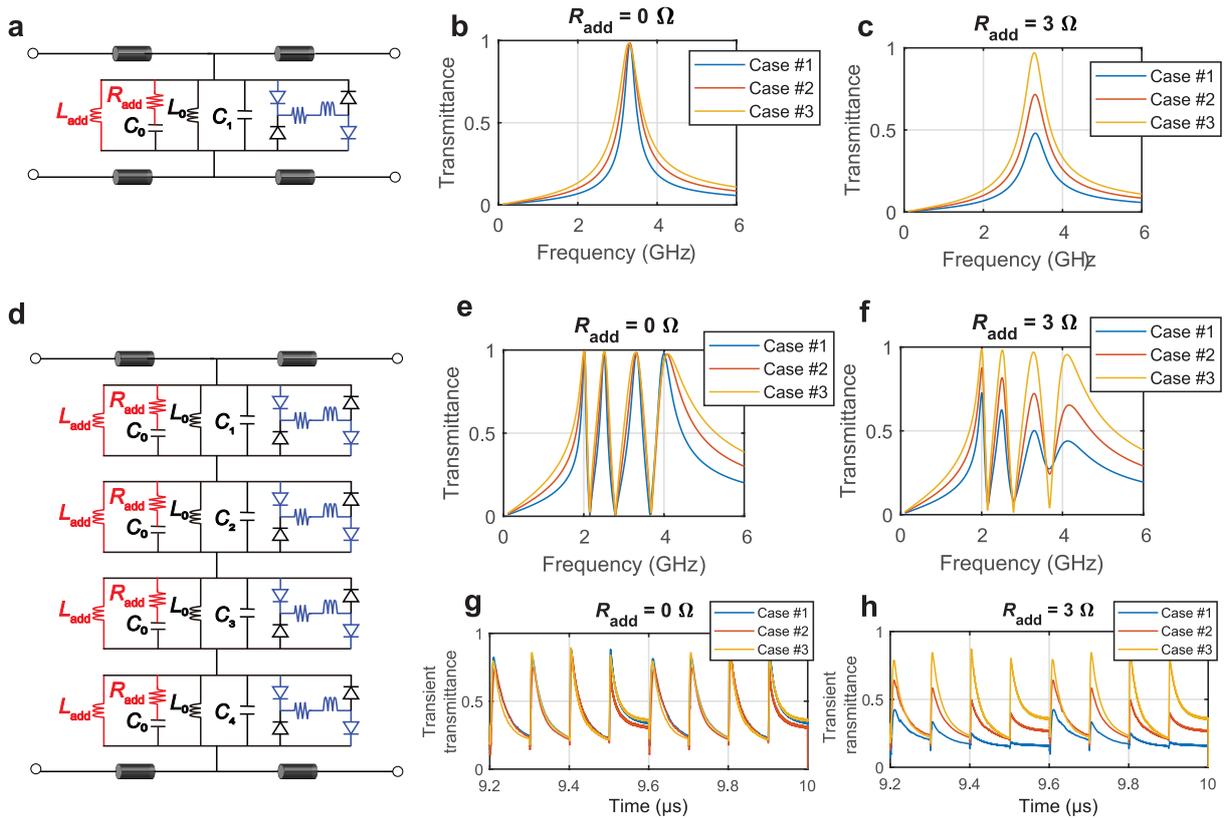

Figure S10: Simplified equivalent transmission line model to represent the metasurface of Fig. 3a for analysing the contribution of the $Q$ factor. (a) Equivalent circuit model connected to the transmission line for the single slit scenario (one unit cell). $C_0$ and $L_0$ represent the capacitive and inductive components of the metallic slit structure. $L_{add}$ and $R_{add}$ represent the inductive and resistive components contributed by the dielectric substrate. The circuit parameters used for the diode and the series inductor and resistor (inside the diode bridge) are the same as those used in Fig. 3. (b, c) Simulated transmittance from linear network analysis for the circuit shown in (a) with varying $Q$ factors while considering (b) zero parasitic resistance $R_{add} = 0\ \Omega$ and (c) nonzero parasitic resistance $R_{add} = 3\ \Omega$ (see Table S7 for variation of $C_1$). (d) Equivalent circuit model connected to the transmission line for the slit scenario (one super cell). (e, f) Simulated transmittance from linear network analysis for the circuit shown in (d) with varying $Q$ factors while considering (e) zero parasitic resistance $R_{add} = 0\ \Omega$, and (f) nonzero parasitic resistance $R_{add} = 3\ \Omega$. (g, h) Nonlinear analysis of transient transmittance for four parallel resonance scenarios with switched frequency. Here, the input power is set to 10 dBm. $f_1, f_2, f_3$ and $f_4$ are 2.0, 2.5, 3.3 and 4.1 GHz, respectively. Simulated transient transmittance with varying $Q$ factor while considering (g) zero parasitic resistance $R_{add} = 0\ \Omega$ and (h) nonzero parasitic resistance $R_{add} = 3\ \Omega$.

Table S7: Circuit values used for Figure S10.

| | Single-resonance scenario | | | |
|---|---|---|---|---|
| | $C_0$ (pF) | $L_0$ (nH) | $L_{add}$ (nH) | $C_1$ (pF) |
| Case #1 | 2.1 | 1 | 1.885 | 0.15 |
| Case #2 | 1.25 | 2 | 1.885 | 0.15 |
| Case #3 | 0.25 | 4 | 1.885 | 0.75 |
| | Four-resonance scenario | | | | | | |
| | $C_0$ (pF) | $L_0$ (nH) | $L_{add}$ (nH) | $C_1$ (pF) | $C_2$ (pF) | $C_3$ (pF) | $C_4$ (pF) |
| Case #1 | 2.1 | 1 | 1.885 | 0.15 | 1.25 | 3.9 | 7.3 |
| Case #2 | 1.25 | 2 | 1.885 | 0.15 | 0.95 | 2.7 | 5 |
| Case #3 | 0.25 | 4 | 1.885 | 0.75 | 1.4 | 2.7 | 4.5 |

**Supplementary Note 5: Supplementary information of measurements related to Fig. 3**

This Supplementary Note provides information related to the measurements associated with Fig. 3. The measurement sample is fabricated using the design parameters of the simulation model explained in Supplementary Note 4. However, some changes are made to deploy the lumped circuit components, as shown in Figure S11, Table S8 and Table S9. The measured frequency-domain transmittance is plotted in Figure S12, where the transmittance for short pulses is much more enhanced than that for CWs at 2.58 GHz and 3.46 GHz. These two frequencies are chosen to perform the measurements associated with Fig. 3.

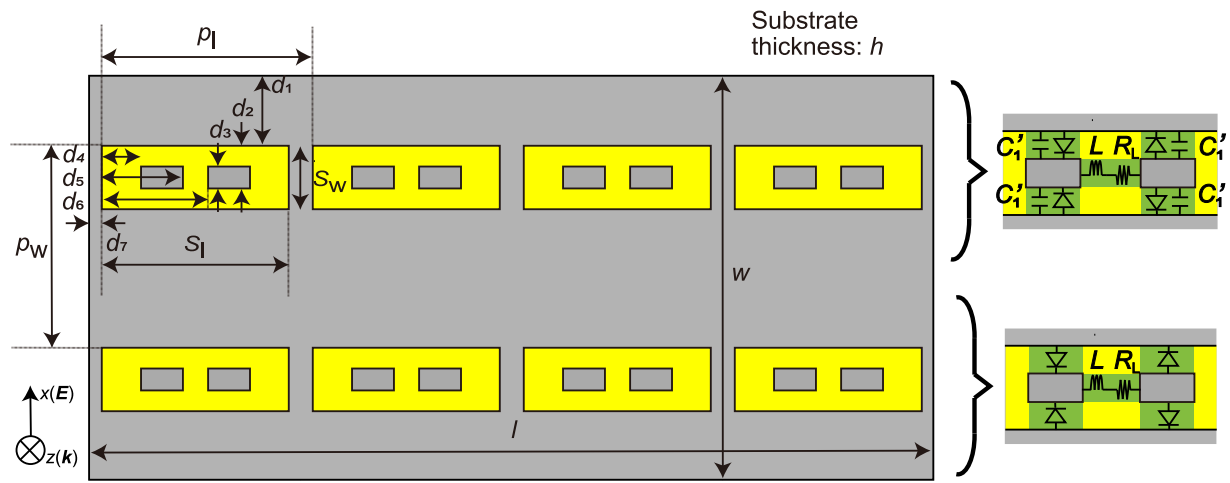

Figure S11: Design of the measurement sample shown in Fig. 3f. The design parameters and circuit values are given in Table S8 and Table S9, respectively. The substrate (yellow) is Rogers3003. In the actual measurement sample, only inductors are used inside diode bridges, as the inductors contain resistive components equal to $R_L$.

Table S8: Design parameters used for the unit cell model drawn in Fig. 3f and Figure S11.

| Parameter | Length [mm] |
| --- | --- |
| $l$ | 72 |
| $w$ | 34 |
| $p_l$ | 18 |
| $p_w$ | 17 |
| $s_l$ | 16 |
| $s_w$ | 5 |
| $h$ | 1.52 |
| $d_1$ | 6 |
| $d_2$ | 3.7 |
| $d_3$ | 1.3 |
| $d_4$ | 4 |
| $d_5$ | 7.5 |
| $d_6$ | 8.5 |
| $d_7$ | 1 |

Table S9: Circuit values used for the unit cell model drawn in Fig. 3f and Figure S11. The value in parentheses represents a self-resonant frequency.

| Parameter | Value |
| --- | --- |
| $L$ | 100 µH (10 MHz) |
| $R_L$ | 5.5 Ω |
| $C_1'$ | 0.3 pF |

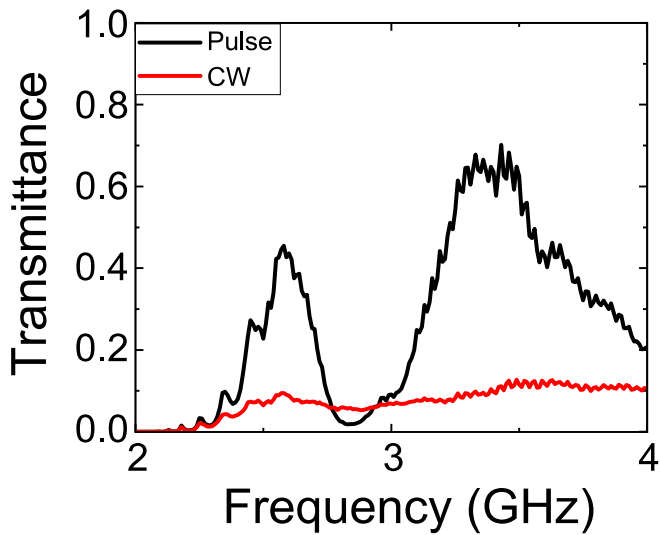

Figure S12: Frequency-domain profile of the transmittance of the measurement sample shown in Fig. 3f.

In addition, another sample is introduced in Figure S13, where the supercells are composed of three types of cells. These cells are designed to operate at different frequencies by introducing additional capacitors, as seen in Figure S13a and Figure S13b. These design parameters are similar to those used for Figure S11, as seen in Table S10 and Table S11. The frequency-domain profiles of this measurement sample are shown in Figure S13c, which supports the operation at three different frequencies, specifically at 2.96, 3.45 and 3.92 GHz. At these frequencies, this measurement sample exhibits limited transient transmittance for CWs in the time domain, as shown in Figure S13d, while repeatedly changing the incoming frequency component increases the transient transmittance, as shown in Figure S13e. Note that the numerically derived transient transmittance for the incident wave using three frequencies, shown in Fig. 3d, is slightly larger than the measured transient transmittance, shown in Figure S13e (see the light blue curve in Fig. 3d). This is because the measurement presented in Figure S13 is limited to a narrow bandwidth due to the cut-off frequencies of the rectangular waveguide used (see Figure S13c). In contrast, the simulation model associated with Fig. 3d does not have such a bandwidth limitation (Figure S4). For this reason, the simulation model is designed to have operating frequencies sufficiently far from each other, which helps avoid interference between different unit cells. Figure S13f shows the transient transmittance for different frequency sequences, which varies only between 0.065 and 0.090. Note again that this limited variation occurs because the measurement sample is not designed to couple unit cells in accordance with the states of other cells, unlike the measurement sample associated with Fig. 4.

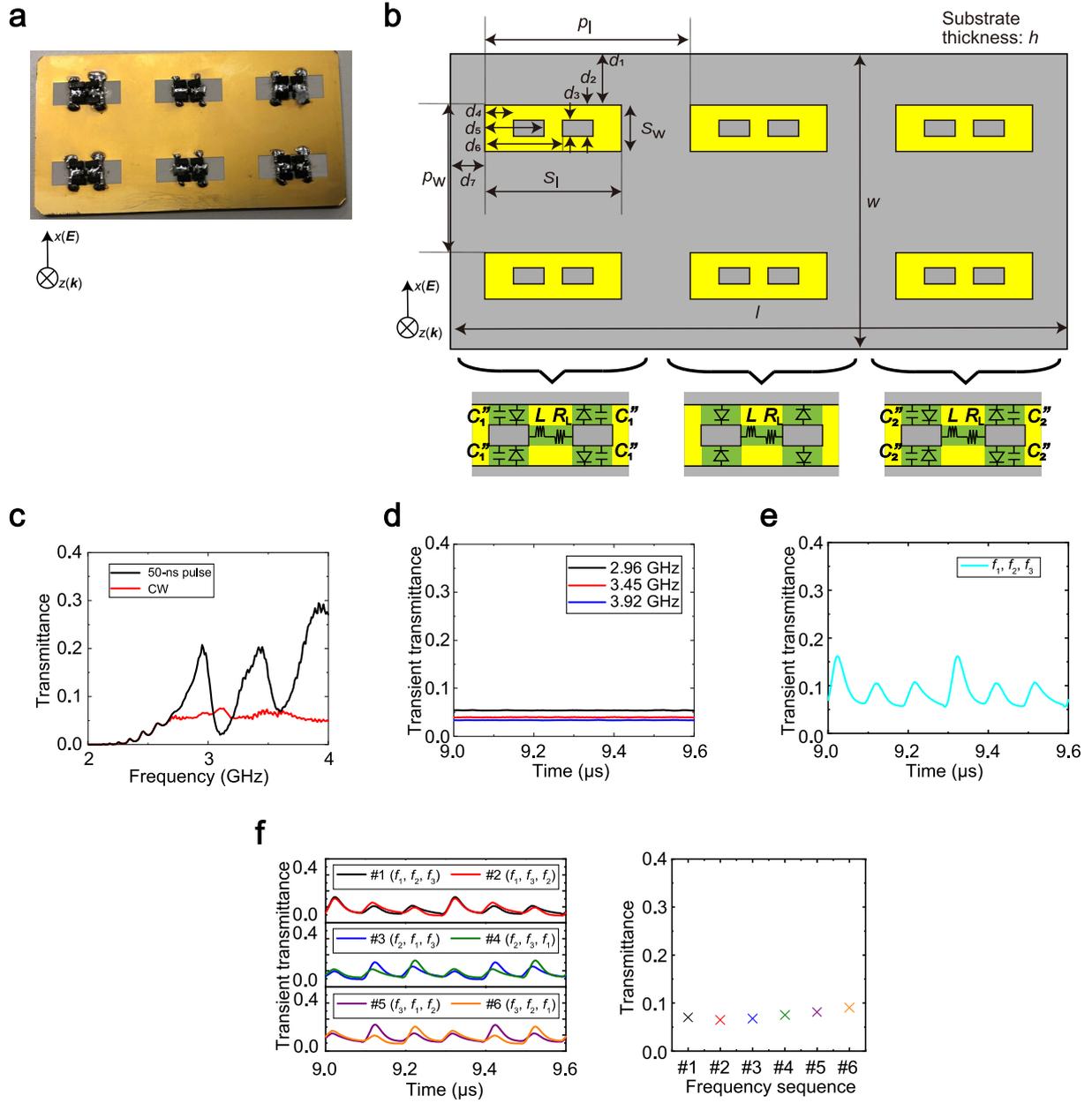

Figure S13: Measurement of the triband waveform-selective metasurface. (a, b) Measurement sample and its dimensions. The design parameters and circuit values used are shown in Table S10 and Table S11. The substrate is Rogers3003. In the actual measurement sample, only inductors are used inside diode bridges, as the inductors contain resistive components equal to $R_L$. (c) Frequency-domain profiles of the transmittance for 50-ns pulses and CWs. $f_1$ = 2.96 GHz, $f_2$ = 3.45 GHz and $f_3$ = 3.92 GHz. Time-domain profiles of the transient transmittances for (d) single-frequency cases and (e) switched-frequency case. (f) Transient transmittance for the switched-frequency case using different frequency sequences.

Table S10: Design parameters used for the measurement sample shown in Figure S13b.

| Parameter | Length [mm] |
|---|---|
| $l$ | 72 |
| $w$ | 34 |
| $p_l$ | 18 |
| $p_w$ | 17 |
| $s_l$ | 16 |
| $s_w$ | 7 |
| $h$ | 1.52 |
| $d_1$ | 5 |
| $d_2$ | 1 |
| $d_3$ | 1 |
| $d_4$ | 4 |
| $d_5$ | 5 |
| $d_6$ | 8 |
| $d_7$ | 2 |
| $d_8$ | 1.5 |
| $d_9$ | 2 |
| $d_{10}$ | 3 |
| $d_{11}$ | 2 |
| $d_{12}$ | 1 |
| $d_{13}$ | 1 |
| $d_{14}$ | 0.5 |
| $d_{15}$ | 1 |
| $d_{16}$ | 14 |

Table S11: Circuit values used for the measurement sample shown in Figure S13b. The value in parentheses represents a self-resonant frequency.

| Parameter | Value |
|---|---|
| $L$ | 100 µH (10 MHz) |
| $R_L$ | 5.5 Ω |
| $C_1"$ | 0.2 pF |
| $C_2"$ | 0.4 pF |

**Supplementary Note 6: Supplementary information of Fig. 4**

The concept of the measurement sample demonstrated in Fig. 4 can be simplified by using two circuit systems, as shown in Figure S14a. In this simplified concept, the left circuit is used to control the right circuit. Based on this concept, the measurement sample depicted in Figure S14b, Figure S14c and Figure S14d that uses supercells composed of only two types of unit cells is fabricated. Basically, these design parameters are the same as those used for the sample associated with Fig. 4 (specifically, see Table S12 and Table S13). Under these circumstances, the measurement sample operates at 2.7 and 3.6 GHz, as shown in Figure S14e. By using these two frequencies, the measurement sample strongly transmits an incoming frequency if the frequency component is changed from 2.7 GHz to 3.6 GHz with 20-dBm input power, as seen in Figure S14f and Figure S14g. This is because the unit cells working at 2.7 GHz are used to vary the JFETs included in the cells operating at 3.6 GHz. Since only two frequencies are used here, the measurement sample exhibits the same transient transmittance in the steady state, as shown in Figure S14h and Figure S14i.

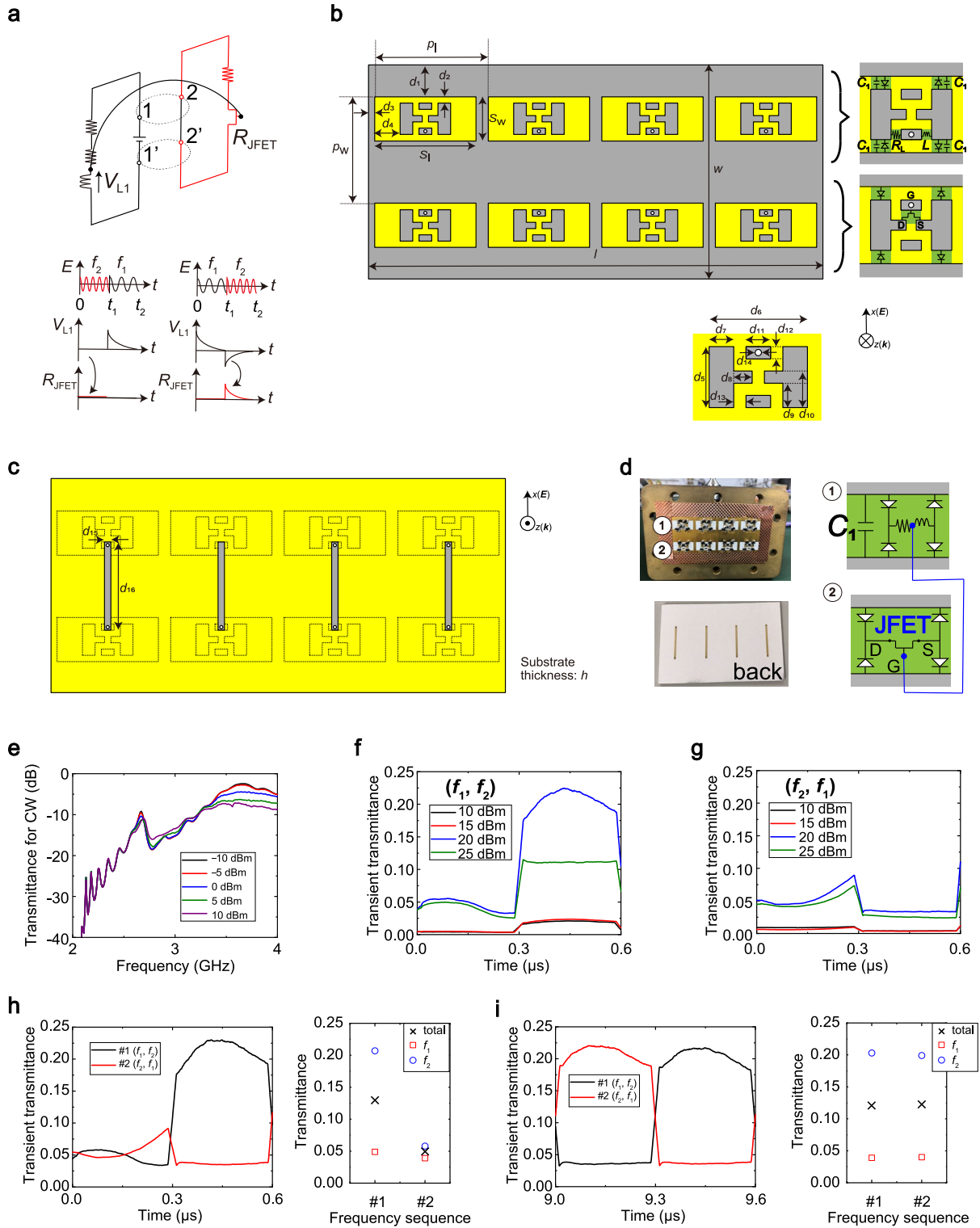

Figure S14: Demonstration of the simplified concept of Fig. 4. (a) Simplified circuit system. Compared to Fig. 4b, only two circuit configurations (i.e., two frequencies) are considered. $V_{L1}$ and $R_{JFET}$ vary in accordance with the incoming frequency. (b) Front and (c) back designs of (d) the measurement sample. The design parameters and circuit values are given in Table S12, Table S13 and Table S14. The substrate is Rogers3003. (e) Frequency-domain profiles of transmittance with various input powers. (f, g) Time-domain profiles of transient transmittance using different frequency sequences. $f_1$ = 2.7 GHz and $f_2$ = 3.6 GHz. The time-

domain transient transmittances (h) during the initial period and (i) in the steady state with 20-dBm input power.

Table S12: Design parameters used for the measurement sample shown in Figure S14b to Figure S14d.

| Parameter | Length [mm] |
|---|---|
| $l$ | 72 |
| $w$ | 34 |
| $p_l$ | 18 |
| $p_w$ | 17 |
| $s_l$ | 16 |
| $s_w$ | 7 |
| $h$ | 1.52 |
| $d_1$ | 5 |
| $d_2$ | 1 |
| $d_3$ | 1 |
| $d_4$ | 4 |
| $d_5$ | 5 |
| $d_6$ | 8 |
| $d_7$ | 2 |
| $d_8$ | 1.5 |
| $d_9$ | 2 |
| $d_{10}$ | 3 |
| $d_{11}$ | 2 |
| $d_{12}$ | 1 |
| $d_{13}$ | 1 |
| $d_{14}$ | 0.5 |
| $d_{15}$ | 1 |
| $d_{16}$ | 14 |

Table S13: Circuit values used for the measurement sample shown in Figure S14b to Figure S14d. The number in parentheses represents a self-resonant frequency.

| Parameter | Value |
|---|---|
| $L$ | 1 mH (2.4 MHz) |
| $R_L$ | 10 kΩ |
| $C_1'$ | 0.2 pF |

Table S14: SPICE parameters used for the JFETs of the measurement sample shown in Figure S14b to Figure S14d. These SPICE parameters are identical to those of a commercial product provided by Toshiba (2SK880-Y).

| Parameter | Value |
|---|---|
| $I_{DSS}$ | 1.2 to 3.0 mA |
| $R_L$ | -0.2 to -1.5 V |
| $C_{iss}$ | 13 pF |
| $C_{rss}$ | 3 pF |

The design parameters and circuit values used for Fig. 4c are provided in Figure S15, Table S15, Table S16 and Table S17. The power dependence of the frequency-domain and time-domain profiles of the sample can be seen in Figure S16. As shown in this figure, the measurement sample is designed to operate near 2.5, 3.3 and 3.9 GHz and that its optimum power level is approximately 20 dBm. In addition, Figure S17 shows how the time-domain profiles change by adjusting the oscillation frequencies. Moreover, Figure S18 provides additional information on the transmittance shown on the right of Fig. 4d.

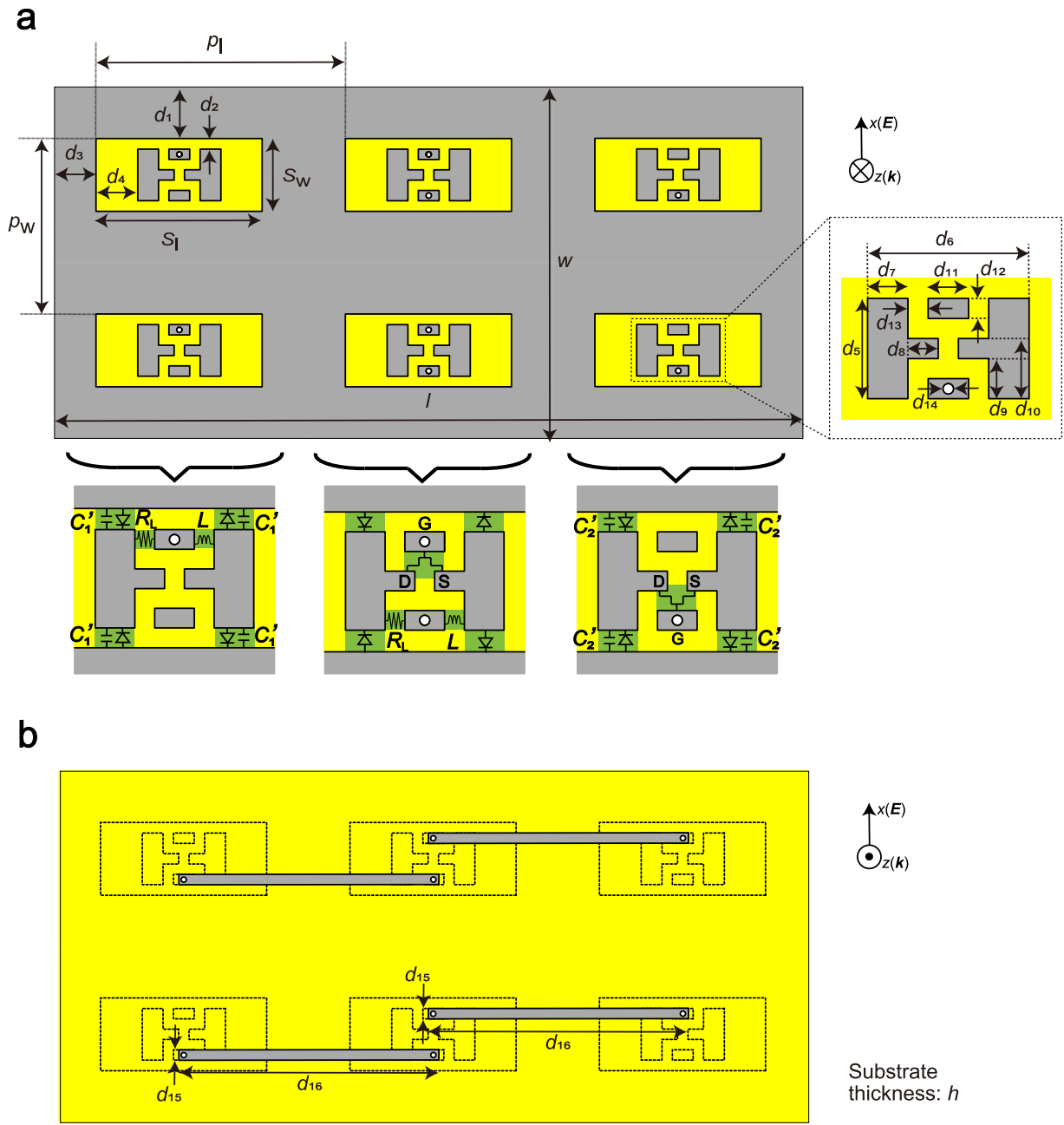

Figure S15: Design of the measurement sample presented in Fig. 4c. (a) Front and (b) back designs. The design parameters and circuit values are given in Table S15, Table S16 and Table S17. The substrate is Rogers3003.

Table S15: Design parameters used for the measurement sample shown in Figure S15 (i.e., Fig. 4c).

| Parameter | Length [mm] |
|---|---|
| $l$ | 72 |
| $w$ | 34 |
| $p_l$ | 24 |
| $p_w$ | 17 |
| $S_l$ | 16 |
| $S_w$ | 7 |
| $h$ | 1.52 |
| $d_1$ | 5 |
| $d_2$ | 1 |
| $d_3$ | 4 |
| $d_4$ | 4 |
| $d_5$ | 5 |
| $d_6$ | 8 |
| $d_7$ | 2 |
| $d_8$ | 1.5 |
| $d_9$ | 2 |
| $d_{10}$ | 3 |
| $d_{11}$ | 2 |
| $d_{12}$ | 1 |
| $d_{13}$ | 1 |
| $d_{14}$ | 0.5 |
| $d_{15}$ | 1 |
| $d_{16}$ | 25 |

Table S16: Circuit values used for the measurement sample shown in Figure S15 (i.e., Fig. 4c). The number in parentheses represents a self-resonant frequency.

| Parameter | Value |
|---|---|
| $L$ | 1 mH (2.4 MHz) |
| $R_L$ | 10 kΩ |
| $C_1'$ | 0.2 pF |
| $C_2'$ | 0.4 pF |

Table S17: SPICE parameters used for the JFETs of the measurement sample shown in Figure S15 (i.e., Fig. 4c). These SPICE parameters are identical to those of a commercial product provided by Toshiba (2SK880-BL).

| Parameter | Value |
|---|---|
| $I_{DSS}$ | 6.0 to 14.0 mA |
| $R_L$ | -0.2 to -1.5 V |
| $C_{iss}$ | 13 pF |
| $C_{rss}$ | 3 pF |

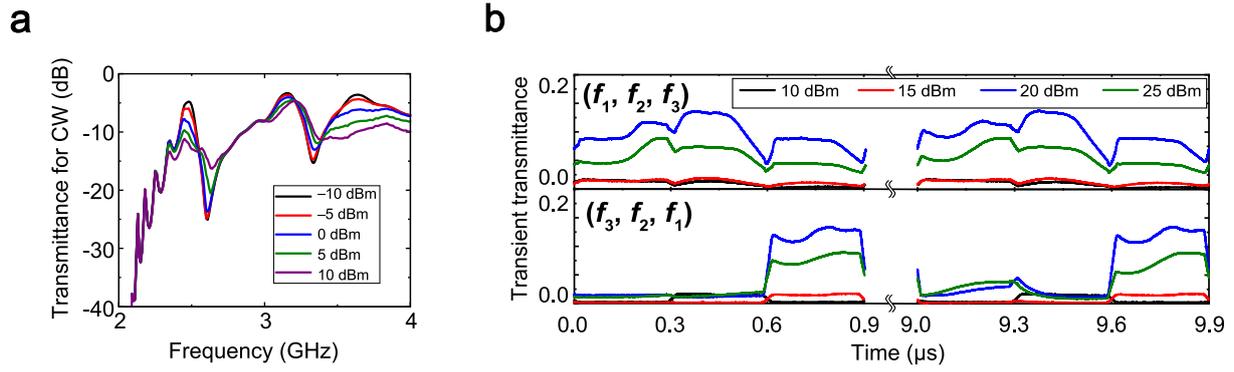

Figure S16: Power dependence of the measurement sample shown in Fig. 4c. (a) Frequency-domain and (b) time-domain profiles. In (b), $f_1$, $f_2$ and $f_3$ were set to 3.3, 3.9 and 2.5 GHz, respectively.

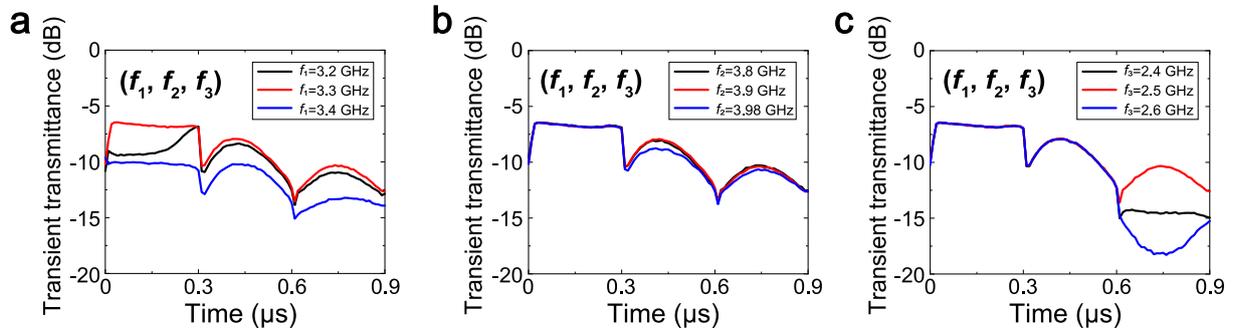

Figure S17: Time-domain profiles of the measurement sample demonstrated in Fig. 4c with adjusted oscillation frequencies. The results with various (a) $f_1$, (b) $f_2$ and (c) $f_3$ values. As the default values, $f_1$, $f_2$ and $f_3$ are set to 3.3, 3.9 and 2.5 GHz, respectively.

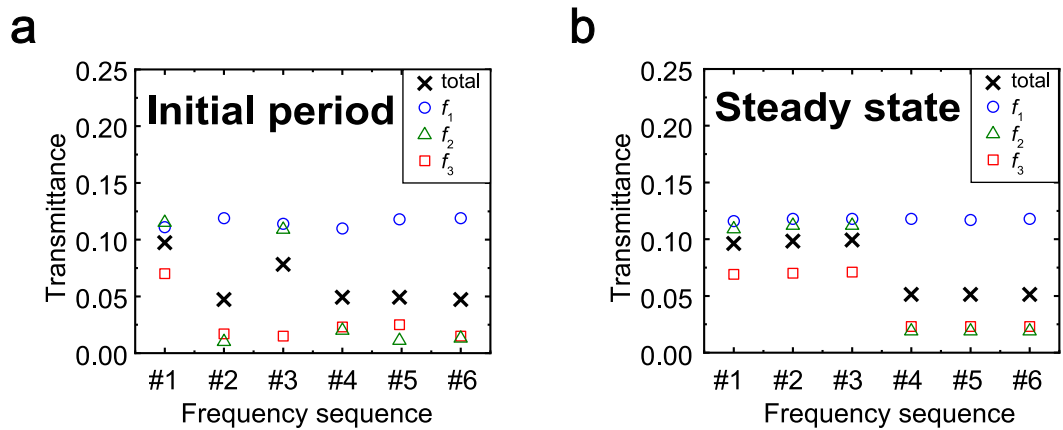

Figure S18: Average transmittance of the measurement sample demonstrated in Fig. 4c for each frequency. $f_1$, $f_2$ and $f_3$ are set to 3.3, 3.9 and 2.5 GHz, respectively.

**Supplementary Note 7: Supplementary information of Fig. 5**

In Fig. 5f, the transmitted energy reduction is calculated including all the frequencies used. The contribution of each frequency component is clarified in Figure S19. Note that the transmitted energy of each frequency component is still compared to that of the signal using sequence #1, as shown in Fig. 5f. Under this circumstance, the transmitted energy of the interference signal increases linearly with I/S. When I/S equals 0 dB (i.e., when the interference magnitude equals the signal magnitude), the transmitted energy of $f_5$ slightly exceeds the transmitted energies of $f_1$, $f_2$ and $f_3$. This is due to the continuous waveform nature of $f_5$ compared to the pulse-based signals in $f_1$, $f_2$ and $f_3$. Nevertheless, as observed in Fig. 5f, the sum of the energies of $f_1$, $f_2$ and $f_3$ is still larger than that of $f_4$ and $f_5$. Particularly, the interference energy exceeds the signal energy of frequency sequence #6 but not the signal energy of frequency sequence #1. This indicates that the presence of adjacent interference results in minimal reduction (less than 3 dB) in the signal energy, ensuring that communication performance can be maintained. However, the influence on the communication performance may vary depending on the level of the interference signal.

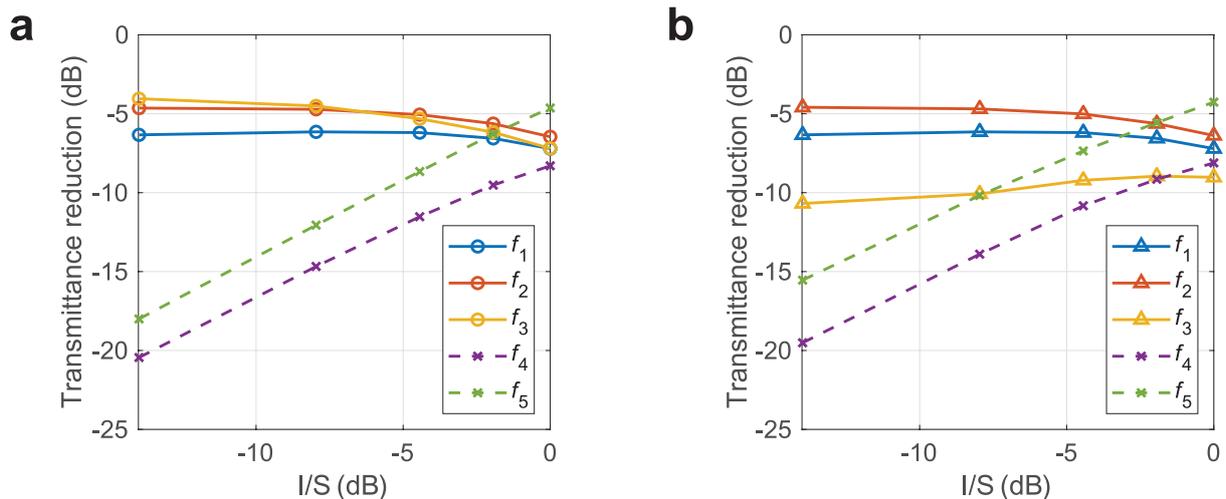

Figure S19: Contribution of each frequency used in Fig. 5f. (a, b) Results for frequency sequences (a) #1 and (b) #6. The result of the interference signal used in Fig. 5f is calculated by the sum of $f_4$ and $f_5$ in (a).

As an additional proof-of-concept experiment, we transmit a binary image using the proposed metasurface configuration and report the obtained results in Figure S20. BPSK modulation with a frequency-hopping carrier is used as explained in the "Modulation Method" of the Methods section (see the diagram block shown in Fig. 5g). Since AWGN is added to the received signals, it is possible to evaluate the retrieved image with variation in the signal-to-noise ratio (S/N). Figure S20 illustrates the differentiation between two frequency sequences as well as the resulting image quality under varying S/N conditions. The retrieved image demonstrates enhanced clarity with fewer errors when the frequency sequence aligns with the embedded metasurface sequence ($f_1$, $f_2$ and $f_3$), and this distinction persists even at higher S/N values (e.g., S/N = 0 dB). These findings align with the BER versus S/N graph presented in Fig. 5i. Importantly, in this communication scenario, both the transmitter and receiver possess knowledge of the frequency sequence used for frequency hopping. Thus, the differentiation arises based on the consistency or inconsistency of the frequency sequence with the one determined or embedded within the metasurface. We also note that the difference between the frequency sequences at the same S/N would be clearer if the gap between the transient transmittances for these frequency sequences are further improved, which leads to an increase in the difference in the BER characteristics in Fig. 5i.

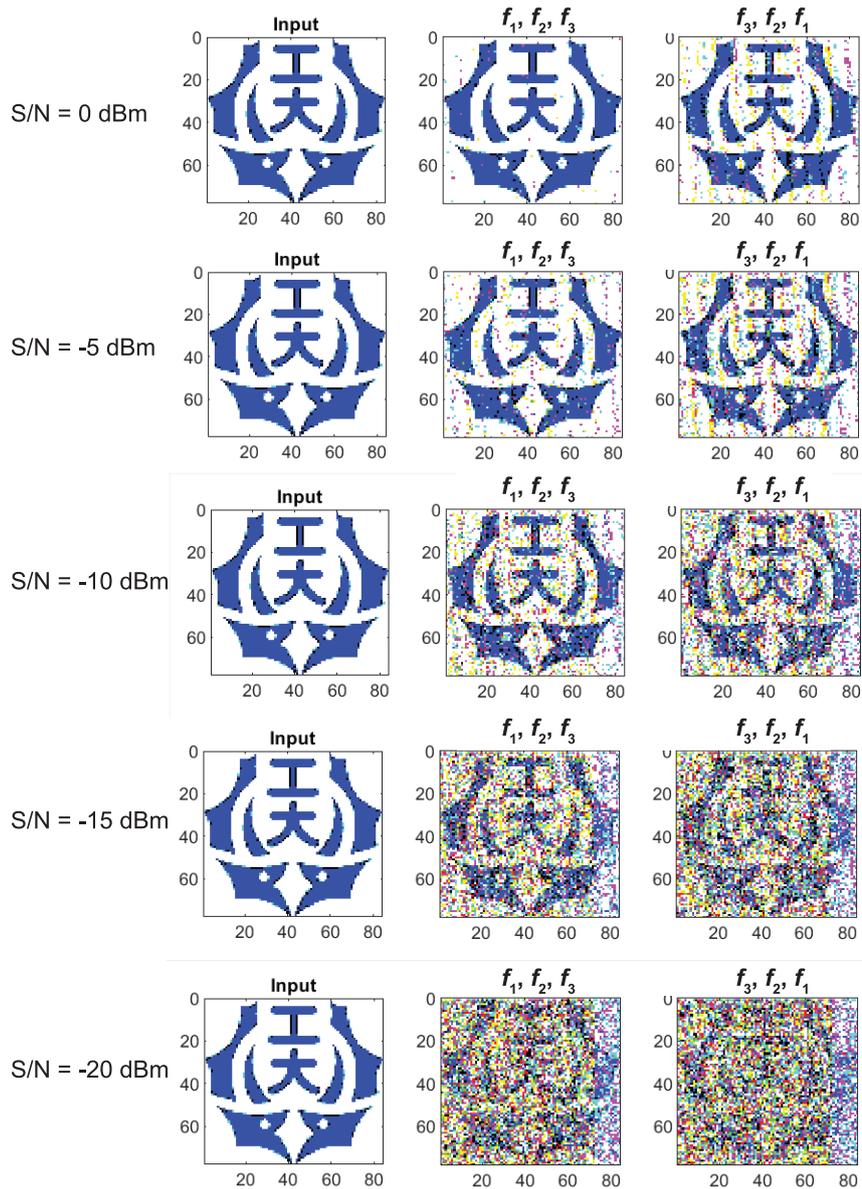

Figure S20: Example of data transmission (logo mark of Nagoya Institute of Technology) using the diagram block shown in Fig. 5g. The S/N is varied from -20 to 0 dBm. Two frequency sequences are used for the carrier, similar to the scenario explained in Fig. 5i.

References

52. Nakasha, T., Phang, S. & Wakatsuchi, H. Pseudo‐waveform‐selective metasurfaces and their limited performance. *Adv Theory Simul* **4**, 2000187 (2021).

53. Wakatsuchi, H. Time-domain filtering of metasurfaces. *Sci Rep* **5**, 16737 (2015).

54. Asano, K., Nakasha, T. & Wakatsuchi, H. Simplified equivalent circuit approach for designing time-domain responses of waveform-selective metasurfaces. *Appl Phys Lett* **116**, 171603 (2020).